\documentclass{article}
\usepackage[table]{xcolor}
\usepackage[export]{adjustbox}

\usepackage{microtype}
\usepackage{booktabs}      % for \toprule, \midrule, \bottomrule
\usepackage{graphicx}      % for \resizebox
\usepackage{pifont}     
\usepackage{booktabs} % for professional tables

\usepackage[most]{tcolorbox}
\tcbuselibrary{listings,breakable}
\definecolor{usercolor}{RGB}{220,240,255}
\definecolor{llmcolor}{RGB}{230,255,230}

\tcbuselibrary{listings,breakable}

\newtcolorbox{userbox}[1][]{
  enhanced,
  breakable,
  listing only,           % <---- 关键！
  colback=usercolor,
  colframe=blue!50!black,
  boxrule=0.8pt,
  arc=3pt,
  left=3pt, right=3pt, top=2pt, bottom=2pt,
  title=#1,
}

\newtcolorbox{llmbox}[1][]{
  enhanced,
  breakable,
  listing only,
  colback=llmcolor,
  colframe=green!60!black,
  boxrule=0.8pt, arc=3pt,
  left=3pt, right=3pt, top=2pt, bottom=2pt,
  title=#1
}

\usepackage{hyperref}
\usepackage{xurl}

\usepackage{multicol}
\usepackage{multirow}
\newcommand{\cmark}{\ding{51}}% 对勾
\newcommand{\xmark}{\ding{55}}% 叉子
\usepackage[accepted]{icml2026}

\usepackage{amsmath}
\usepackage{amssymb}
\usepackage{mathtools}
\usepackage{amsthm}

\usepackage{graphicx}
\usepackage{subcaption}
\usepackage[capitalize,noabbrev]{cleveref}

\theoremstyle{plain}

\theoremstyle{definition}

\theoremstyle{remark}

\usepackage[textsize=tiny]{todonotes}

\begin{document}
\chead{\textbf{One Tool Is Enough: Reinforcement Learning of LLM Agents for Repository-Level Code Navigation}}
\twocolumn[
\icmltitle{One Tool Is Enough: Reinforcement Learning of LLM Agents for Repository-Level Code Navigation}


% You can specify symbols, they are numbered in order.
\icmlsetsymbol{equal}{*}

\begin{icmlauthorlist}
\icmlauthor{Zhaoxi Zhang}{pku_lab,pku,equal}
\icmlauthor{Yitong Duan}{zgca}
\icmlauthor{Yanzhi Zhang}{zgca}
\icmlauthor{Yiming Xu}{pku}
\icmlauthor{Zhixiang Wang}{pku_lab,pku}
\icmlauthor{Kun Liang}{pku_lab,pku}
\icmlauthor{Yang (Weikang) Li}{baidu}
\icmlauthor{Jiahui Liang}{baidu}
\icmlauthor{Deguo Xia}{baidu}
\icmlauthor{Jizhou Huang}{baidu}
\icmlauthor{Jiyan He}{zgca}
\icmlauthor{Shuxin Zheng}{zgca}
\icmlauthor{Yunfang Wu}{pku_lab,pku}

\end{icmlauthorlist}
\icmlaffiliation{pku_lab}{National Key Laboratory for Multimedia Information Processing, Peking University}
\icmlaffiliation{pku}{School of Computer Science, Peking University}
\icmlaffiliation{zgca}{Zhongguancun Institute of Artificial Intelligence (ZGCI)}
\icmlaffiliation{baidu}{Baidu Inc}

\icmlcorrespondingauthor{Yitong Duan}{duanyitong@\allowbreak zgci.ac.\allowbreak cn}
\icmlcorrespondingauthor{Yunfang Wu}{wuyf@\allowbreak pku.\allowbreak edu.\allowbreak cn}

\icmlkeywords{Machine Learning, ICML}

\vskip 0.3in
]



\printAffiliationsAndNotice{*Work done during internship at ZGCI.}

\begin{abstract}
Locating files and functions requiring modification in large software repositories is challenging due to their scale and structural complexity. Existing LLM-based methods typically treat this as a repository-level retrieval task and rely on multiple auxiliary tools, which often overlook code execution logic and complicate model control. We propose \textbf{RepoNavigator}, an LLM agent equipped with \textbf{a single execution-aware tool}: jumping to the definition of an invoked symbol. This unified design reflects the actual flow of code execution while simplifying tool manipulation. RepoNavigator is \textbf{trained end-to-end via Reinforcement Learning (RL)} directly from a base pretrained model, without relying on closed-source distillation. Experiments demonstrate that RL-trained RepoNavigator achieves state-of-the-art performance, with the 7B model outperforming 14B baselines, the 14B model surpassing 32B competitors, and the 32B model exceeding closed-source models such as GPT-5 on most metrics. These results confirm that integrating \textbf{a single, structurally grounded tool with RL training} provides an efficient and scalable solution for repository-level issue localization. 
\end{abstract}
\section{Introduction}

With the rapid advancement of Large Language Models (LLMs) 
\citep{liu2024deepseek,team2024qwen2,yang2025qwen3}, equipping LLMs with pre-built tools to form LLM agents has become a common paradigm for 
expanding their capabilities \cite{shen2024llm,yuan2024easytool,lu2024toolsandbox}. 
In the domain of software engineering (SWE), although LLM agents 
can effectively handle simple programming tasks 
\cite{hui2024qwen2,guo2024deepseek}, their ability to operate on large-scale software repositories remains limited. % 这里也是一样的问题，和是否开源有什么关系？闭源也适用吧，只是我们能获得开源数据，所以用开源软件形成的数据集做训练和测试。
{SWE-bench} \citep{jimenez2023swe} currently serves as the most 
comprehensive benchmark for evaluating whether LLMs can resolve real-world 
GitHub issues. All pretrained LLMs cannot process the whole repository directly due to context limits \cite{yang2024swe}. % 这里或许加个引用？（如果还有空的话）
While {SWE-Agent} \citep{jimenez2023swe} provides moderate gains, 
it remains far from enabling robust repository-level reasoning.

\begin{figure}[t]
    \centering
    \includegraphics[width=1\linewidth]{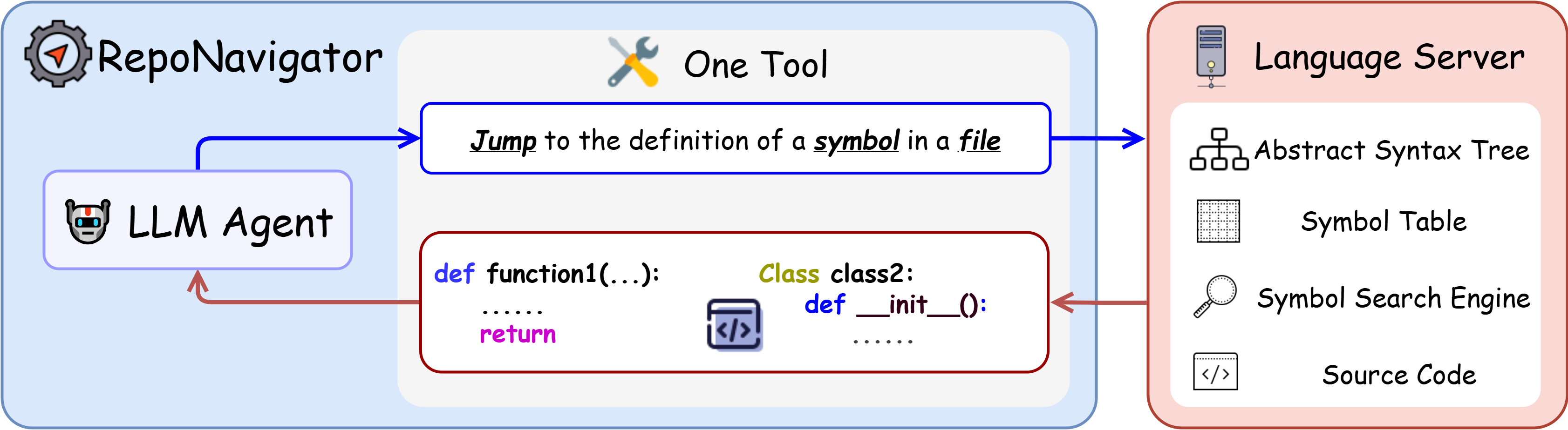}
    \caption{
    Illustration of an LLM navigating through a code repository. The LLM is equipped with a single yet powerful tool: \texttt{jump}, implemented via a language server.}
    \label{fig:agent_pipeline}
\end{figure}
% 你这个图不需要再正文中提一下的吗？它和正文的关系是什么？我应该阅读到什么地方的时候看这个图？会使读者困惑

Most existing agents rely on test-time scaling applied directly to pretrained LLMs \cite{liu2024dynamic,chen-etal-2025-locagent,schmidgall2025agent}.
In SWE tasks% 前面已经有过software engineering (SWE)了，这里不用两个都出现了吧，只写全称或者简称就行了，后面也一样。
, tools are essential rather than optional: real-world repositories are far larger than the context window of current LLMs, making it impossible to process an entire codebase in a single forward pass. % 和摘要部分我提出的一样，可以加个参考文献
Agents must therefore iteratively invoke tools to retrieve partial information from the repository and interleave natural-language reasoning with tool calls.

Mainstream LLM agents \cite{chen-etal-2025-locagent,liu2025codexgraph,xiang2025promptsculptor,wang2023promptagent,chen2024comm} are rarely exposed to such agentic interaction patterns during pretraining and typically acquire tool usage only through few-shot prompting, which is insufficient for learning complex multi-step tool-chaining behaviors. Moreover, because tool definition spaces are effectively unbounded, pretrained models cannot fully internalize their semantics without post-training.
To mitigate these issues, post-training paradigms such as Supervised Finetuning (SFT) \cite{ma2025toolintegratedreinforcementlearningrepo} and Reinforcement Learning with Verifiable Rewards (RLVR) \cite{yu2025dapo,yue2025vapo} have been applied, with promising results in domains including retrieval agents \cite{jin2025search}, and GUI agents \cite{hong2024cogagent}

Directly training an agent to fix software issues, however, remains 
difficult. A single bug often admits multiple valid patches, making 
string-level evaluation unreliable. % 这里的string-level具体指的是什么？我觉得可以简单解释一下（如果这是这个领域的很基础的知识的话，当我没说）
% zzx: 这块说了docker需要很多cpu资源，定位的评估几乎不需要。感觉没法说具体多少资源...
The only precise evaluation method 
requires executing candidate patches inside a dedicated Docker environment 
for each repository \cite{deepswe2025}, which is prohibitively expensive in terms of CPU resources for supervised training. 
To make training more tractable, we adopt a simplified yet widely 
generalizable assignment: \textbf{issue localization}. 
Prior work shows that a software issue becomes substantially easier to 
resolve once the relevant functions and files are correctly identified 
\cite{chen-etal-2025-locagent,ma2025toolintegratedreinforcementlearningrepo,xia2024agentless,jiang2025cosil}. 
Since modern software repositories contain a significant amount 
of code—far beyond any LLM’s context window—localization drastically 
reduces the search space and improves downstream solvability. 
Crucially, localization outputs a discrete set of paths, enabling 
verifiable, string-level evaluation that is compatible with scalable 
training frameworks such as SFT and RLVR (string-level evaluation requires little resource).

Existing localization agents 
\cite{ma2025toolintegratedreinforcementlearningrepo,chen-etal-2025-locagent,SWESwiss2025} 
typically rely on multiple tools, including \texttt{SearchClass}, 
\texttt{SearchMethods}, and \texttt{GetImports}. 
Although effective to some extent, these tools consider high-level abstractions (classes, functions, etc) of programming languages, which do not reflect how code 
actually executes. High-level abstractions, such as classes or inheritance, 
disappear after compilation, leaving only sequential execution and \texttt{jump} 
operations. Since modern LLMs already excel at capturing sequential 
dependencies, we focus on enhancing their ability to \texttt{jump} across 
the repository—that is, to follow and inspect the source definition of 
symbols as they appear in execution. To this end, we introduce a single, 
structurally grounded tool: \texttt{jump}, which retrieves the precise 
definition of a given symbol. Our agent invokes this single tool to navigate in the repository, as depicted in Fig.~\ref{fig:agent_pipeline}. Details of this tool are provided in 
Sec.~\ref{sec:tool_definition}.

Our main contributions are threefold:  
(1) We propose the first repo-level localization agent trained using reinforcement learning directly from the pretrained model, without relying on distillation from a closed-source model.
(2) We design a repository navigation agent that operates by performing 
realistic \texttt{jump} operations aligned with actual execution semantics.  
(3) We demonstrate that one unified tool 
significantly improves efficiency and controllability compared to 
multi-tool pipelines.

\section{Related Works}
\label{sec:related_works}
\subsection{Agentic Training}
LLM agents are promising methods to equip models with complex tools while reasoning \cite{li2024personal,huang2024understanding,guo2024large}. However, because most pretrained LLMs are trained on texts only, and developers can define any tools, most tools are out-of-domain (OOD) for LLMs. Even for the most powerful models, failures often happen when calling the newly defined tools due to a wrong calling format or failed parameter parsing. Thus, training an LLM to master a newly defined tool is critical for LLM agents. Intuitively, the tool-calling trajectories can be generated by a more powerful LLM, and such trajectories can be used to train a student model via supervised finetuning (SFT) \cite{chen-etal-2025-locagent}. However, this pipeline requires a stronger teacher model that has the capability to master the tool. Recently, more methods have emerged with no teacher model required. Rejected-sampled finetuning (RFT) \cite{ahn2024large} utilizes generated trajectories of the agent itself via multiple rollouts. Agentic RL \cite{jin2025search} is an on-policy RLVR method requiring only the result for verifying trajectories. Such training methods yield remarkable results when the tools are search engines \cite{jin2025search}, Python executers \cite{jimenez2023swe}, calculators \cite{yan2025mathagent}, and visual models \cite{gupta2023visual}.

\begin{figure*}
    \centering
    \includegraphics[width=1\linewidth]{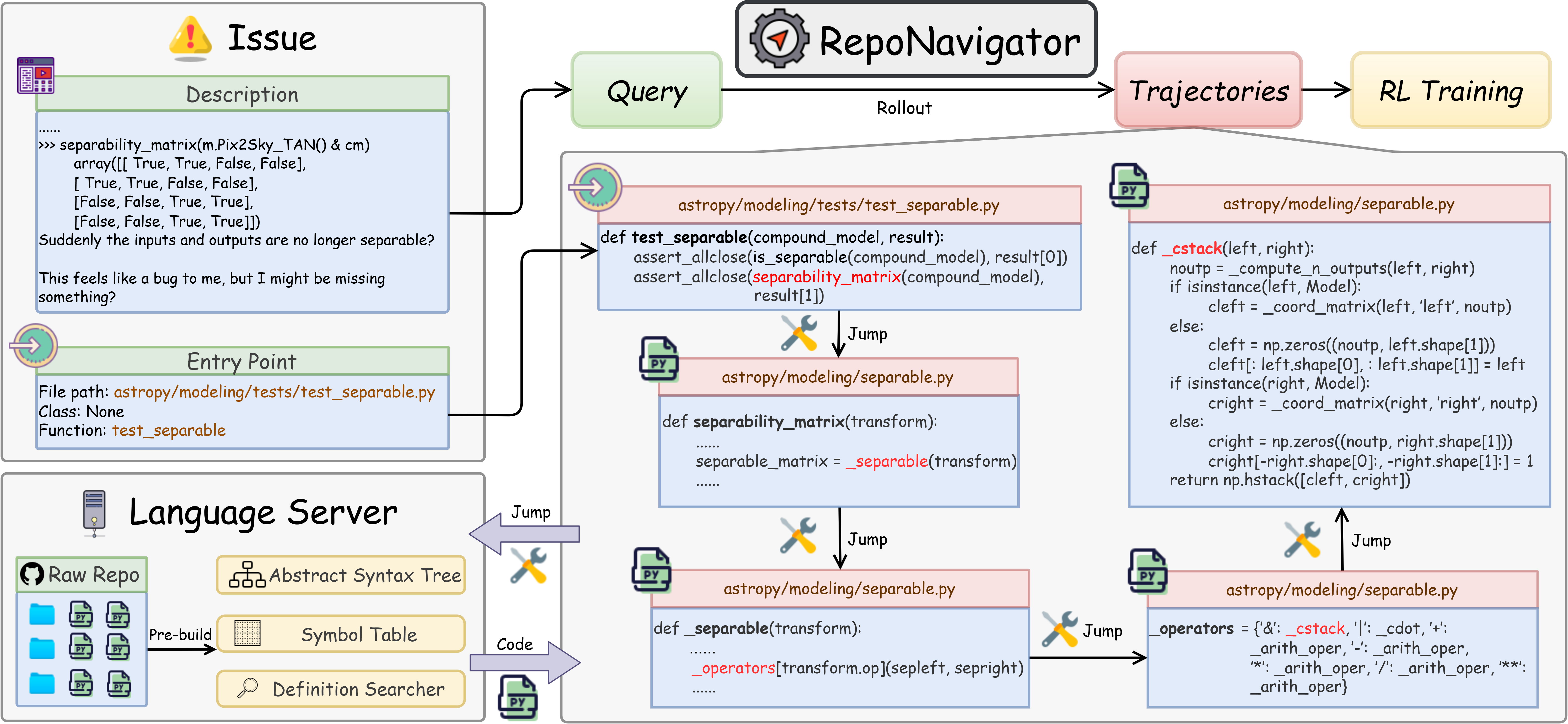}    \caption{Overview of our RepoNavigator. During the rollout phase, the agent can invoke the \texttt{jump} tool, and the language server will return the definition code of the symbol. This process is trained by reinforcement learning.}
    \label{fig:placeholder}
\end{figure*}
\subsection{Software Engineering Agents}

The introduction of {SWE-bench} \cite{jimenez2023swe,yang2024swe} has motivated a range of agentic pipelines for software engineering tasks. % 这里也一样，全称和简称用一个就行了吧
Among them, {SWE-agent} \citep{yang2024sweagent} and {OpenHands} \citep{wang2025openhands} are widely adopted frameworks that equip agents with tools for interacting with computing environments (terminals in Linux). Workflow-based methods such as {Agentless} \cite{xia2024agentless} decompose issue resolution into localization, repair, and validation subproblems. 

While {DeepSWE} \cite{deepswe2025} and {SWE-Swiss} \cite{SWESwiss2025} achieve high performance through reinforcement learning, end-to-end training remains computationally expensive due to the overhead of executing Docker containers for patch evaluation. Consequently, issue localization has emerged as a lightweight alternative, focusing on identifying faulty files or functions. Empirical evidence suggests that superior localization significantly improves overall issue resolution efficiency \cite{chen-etal-2025-locagent,xia2024agentless,yu2025orcalocallmagentframework,ma2025toolintegratedreinforcementlearningrepo}.

Recent localization agents include: {LocAgent} \citep{chen-etal-2025-locagent}, {CoSIL} \cite{jiang2025cosil}, and {GraphLocator} \cite{liu2025graphlocator}, which model codebases as graphs and integrate them into LLMs, and {OrcaLoca} \citep{yu2025orcalocallmagentframework}, which enhances efficiency through priority scheduling, action decomposition, and context pruning. From an open-source perspective, {RepoSearcher} \citep{ma2025toolintegratedreinforcementlearningrepo}, trained with distillation and RL on the Qwen model family \citep{team2024qwen2}, represents a notable advancement.

Nevertheless, prior agents overlook the structural relations within repositories—where modules, classes, and functions are cross-referenced across files—and typically rely on multiple search tools for symbol definition retrieval, amplifying error propagation (see Sec.~\ref{sec:method}). In contrast, we employ a single execution-logic-focused tool, reducing usage complexity. Finally, our approach constitutes the first localization agent trained directly from pretrained models, without relying on distillation-based supervised finetuning, a crucial stage in both RepoSearcher \cite{ma2025toolintegratedreinforcementlearningrepo} and LocAgent \cite{chen-etal-2025-locagent}.

\section{Method}
\label{sec:method}
We present \textbf{RepoNavigator}, a reinforcement-learning agent for repository-level issue localization. The method consists of three components: (1) a unified tool to retrieve the definition of any symbols in a given file, (2) a reasoning--action agent loop that alternates between natural-language reasoning and tool invocation, and (3) a GRPO-based RL algorithm for optimizing long-horizon tool-augmented trajectories.

\subsection{Problem Formulation}
Given a repository $\mathcal{R}=\{f_1,\dots,f_N\}$ and an issue description $q$, the goal is to output relevant code regions $Y^*=\{(f_i,g_{i,j})\}$, where $g_{i,j}$ denotes a function or code span in file $f_i$. At each step $t$, the agent produces an optional reasoning step $r_t$, a tool call $a_t$, and receives the observation $o_t$, forming a trajectory $\tau=\{(r_t,a_t,o_t)\}_{t=1}^T$. After termination, a final prediction $\hat{Y}$ is scored by a reward $R(\hat{Y},Y^*)$. The objective is $\max_\theta \mathbb{E}_{\tau\sim\pi_\theta}[R(\tau)]$.

\subsection{Agent Architecture}
RepoNavigator uses a \emph{single-tool} design to avoid multi-tool orchestration overhead. At each step, the policy $\pi_\theta$ decides whether to continue reasoning or to emit a JSON-formatted tool call, while a symbol and its corresponding file are parsed to the tool.
The agent receives structured observations (code snippets or error messages), then continues reasoning until termination. The loop is \textit{reason → act → observe} \cite{yao2022react}. Given history $h_t=(q,o_{1:t-1},a_{1:t-1})$, the agent samples either a natural-language reasoning step $r_t\sim\pi_\theta(\cdot|h_t)$ or a structured tool call $a_t\sim\pi_\theta(\cdot|h_t)$. Tool calls must satisfy a JSON grammar enforced via constrained decoding. The loop continues until the agent predicts its final locations $\hat{Y}$.

\subsection{Jump: Symbol Resolution}
\label{sec:tool_definition}

Language servers resolve the definition of a symbol via a deterministic static analysis pipeline that approximates runtime name-binding semantics. Given a symbol occurrence \( s \) at source location \( \ell \), the resolution process computes a mapping from the source location to the definition location:
\begin{equation}
    \mathcal{R}(s, \ell) \rightarrow \{(f_i, p_i)\},
\end{equation}
where each pair \( (f_i, p_i) \) denotes a file path and source position corresponding to a valid definition of \( s \).
In practice, we provide \texttt{file\_path} and \texttt{symbol} to locate \( \ell \); when multiple symbols share the same name within a snippet, an additional \texttt{index} is used to disambiguate the targets.

\paragraph{Syntactic and Scope Analysis}
The source file is first parsed into an abstract syntax tree (AST). The syntactic role of \( s \) (e.g., name or attribute access) determines the resolution strategy.
For name symbols, candidate definitions are searched along the lexical scope chain
\begin{equation}
    \mathcal{S} = \{\text{local}, \text{enclosing}, \text{module}, \text{builtins}\},
\end{equation}
following the LEGB rule (for languages like Python, C++, etc.), with each scope maintaining a symbol table.

\paragraph{Type-Based Attribute Resolution}
For attribute expressions \( a.b \), the language server infers a (possibly union-valued) type for the receiver \( a \) using annotations, control-flow analysis, function return types, and stub files.
Member lookup is then performed over the method resolution order (MRO), denoted as:
\begin{equation}
\mathrm{resolve}(a.b) = \bigcup_{t \in T(a)} \mathrm{lookup}(b, \mathrm{MRO}(t)).
\end{equation}

\paragraph{Import Dependency Graph}
Cross-file resolution relies on an import dependency graph that statically emulates the module loading semantics. Import statements introduce bindings from local names to exported symbols of target modules, including re-exports and \texttt{\_\_all\_\_}-based filtering, and resolution may traverse multiple modules before reaching a concrete definition.

\paragraph{Role of Jump}
The jump operation is not an oracle that directly reveals task-relevant code. Rather, it abstracts execution-aware navigation already provided by modern IDE language servers for almost all programming languages. Consequently, the agent still faces a long-horizon decision problem: selecting which jump sequences to execute and when to stop exploration. We address this by training the agent with reinforcement learning in the next section.

% \subsection{Reasoning-action Loop}
% Given history $h_t=(q,o_{1:t-1},a_{1:t-1})$, the agent samples either a natural-language reasoning step $r_t\sim\pi_\theta(\cdot|h_t)$ or a structured tool call $a_t\sim\pi_\theta(\cdot|h_t)$. Tool calls must satisfy a JSON grammar enforced via constrained decoding. The loop continues until the agent predicts its final locations $\hat{Y}$.

\subsection{Reinforcement Learning}
We apply reinforcement learning with verifiable rewards to train the agent directly from the pretrained model. We do not require a more powerful teacher model for distillation as a warmup, which is the technique of \citep{ma2025toolintegratedreinforcementlearningrepo}. In practice, we apply Group Relative Policy Optimization (GRPO) \cite{liu2024deepseek}, which has the loss function:

\begin{multline}
\label{eq:grpo}
    \mathcal{L}^{\text{GRPO}}(\theta) = 
    \mathbb{E}_{(s_t, a_t) \sim \pi_{\theta_{\text{old}}}} [ 
    \frac{\pi_{\theta}(a_t|s_t)}{\pi_{\theta_{\text{old}}}(a_t|s_t)} \hat{A}_t \\
    - \left. \beta \, D_{\text{KL}}\left( \pi_{\theta_{\text{old}}}(\cdot|s_t) \| \pi_{\theta}(\cdot|s_t) \right) 
    \right]
\end{multline}

where the first term is the standard policy gradient objective with an estimated advantage function \(\hat{A}_t\), which promotes actions that lead to higher-than-expected returns. The second term is a Kullback-Leibler (KL) divergence penalty, scaled by a coefficient \(\beta\), which acts as a trust region, preventing the updated policy \(\pi_{\theta}\) from moving too far from the previous policy \(\pi_{\theta_{\text{old}}}\). This formulation ensures stable and consistent policy improvement by balancing reward maximization with behavioral consistency. The advantage \(\hat{A}_t\) is calculated in the standard GRPO method, and Algorithm~\ref{alg:reponavigator} presents the full RL training process of RepoNavigator.

The reward of the GRPO process is calculated as:
\begin{equation}
R(\hat{Y}, Y^*, \tau)
= \lambda \cdot \mathrm{DICE}(\hat{Y}, Y^*) 
+ (1-\lambda)\cdot\mathrm{S}(\tau)
\end{equation}
where \(\lambda\) is heuristically set to \(0.5\) to balance the tool reward and the outcome reward (we prove the insensitivity of our method to \(\lambda\) in Appendix~\ref{sec:hyperparameter}.). For the predicted set $\hat{Y}$, and the groundtruth set $Y^*$
\begin{equation}
    \text{DICE}(\hat{Y}, Y^*) = \frac{2\times|\hat{Y}\cap  Y^*|}{|\hat{Y}|+| Y^*|}
\end{equation}
and $S(\tau)$ is the success rate of tool-calling extracted from $\tau$. We consider the tool call to be failed when the format is incorrect, or the symbol parsed does not exist, or for any other reason that results in unexpected tool termination. 
% \begin{algorithm}[H]

% \KwData{ Repo $\mathcal{R}$, issue $q$, initial policy $\pi_\theta$}
% \KwResult{$\hat{Y}$ maximizing $\mathbb{E}_{\tau\sim\pi_\theta}[R(\tau)]$}

% $o_0 \leftarrow q,\;\tau\leftarrow\emptyset$\;

% \While{not terminated}{
%   $h_t \leftarrow (q,o_{1:t-1},a_{1:t-1})$\;
  
%   % 可以在此产生任意 reasoning 内容，但不显式建模
%   (Optionally generate intermediate reasoning)\;
  
%   $x_t \sim \pi_\theta(\cdot|h_t)$\;
  
%   \eIf{$x_t$ is \textbf{tool-call}}{
%     $a_t \leftarrow x_t$\;
    
%     $o_t = T(a_t, \mathcal{R})$\;
    
%     $\tau \leftarrow \tau \cup \{(a_t,o_t)\}$\;
%   }{
%     % final answer
%     $\hat{Y} \leftarrow x_t$\;
    
%     \textbf{terminate}\;
%   }
% }

% $R \leftarrow R(\hat{Y},Y^*, \tau)$\;

% \BlankLine

% % Estimate $\hat{A}_t$ for $(s_t,a_t)\!\in\!\tau$\;
% % \BlankLine
% Estimate $\mathcal{L}^{\text{GRPO}}(\theta)$

% \caption{Pseudo code of the RL process of RepoNavigator}
% \end{algorithm}

\begin{table*}[!t]

\centering
% \vspace{-4 pt}
\caption{Comparison of different agent pipelines on function-level and file-level metrics on SWE-bench\_Verified. We use Qwen2.5-Instruct series as our base model. \textbf{Bold numbers} denote the best performance among same-size models; \underline{underline numbers} denote the best training-free performance among same-size models; \colorbox{yellow!20}{yellow background} illustrates training-free RepoNavigator; \colorbox{blue!10}{blue background} illustrates RepoNavigator trained with GRPO.}
\label{tab:agent_comparison_verified}
\small
\setlength{\tabcolsep}{5pt}
\renewcommand{\arraystretch}{1.2}
\begin{tabular}{cc|cccc|cccc}

\toprule
% \multirow{2}{*}{\textbf{Model}} & 
\multirow{2}{*}{\textbf{Agent Pipeline}} & 
\multirow{2}{*}{\textbf{Model}} & 
\multicolumn{4}{c|}{\textbf{Function-level}} &
\multicolumn{4}{c}{\textbf{File-level}} \\
\cmidrule(lr){3-6} \cmidrule(lr){7-10}
 & & {Recall} & {Precision} & {Sample-F1} & {IoU} 
& {Recall} & {Precision} & {Sample-F1} & {IoU} \\
\midrule
\midrule
& \multicolumn{7}{c}{\textbf{\textit{closed-source Models}}} \\
\midrule
  % LocAgent & Claude3.5-Sonnet & 17.62 & 11.71 & 12.71 & 10.31 & 60.96 & 34.88 & 40.67 & 33.33\\
  RepoSearcher & Claude3.7-Sonnet & \textbf{66.80} & 19.90 & 28.30 & 17.89 & \textbf{89.71} & 21.04 & 33.15 & 20.67\\
  RepoNavigator & Claude3.7-Sonnet & 31.03 & 34.43 & 31.72 & 30.22 & 72.26 & 75.95 & 73.01 & 71.37\\
  RepoNavigator & GPT5-chat & 30.42 & 34.56 & 31.17 & 29.67 & 58.17 & 61.87 & 58.88 & 57.33\\
  RepoNavigator & Claude4.5-Sonnet & 43.97 & \textbf{45.76} & \textbf{43.62} & \textbf{41.31} & 80.68 & \textbf{81.92} & \textbf{79.94} & \textbf{77.49}\\
\midrule
\midrule
& \multicolumn{7}{c}{\textbf{\textit{Qwen2.5-7B}}} \\
\midrule
  Agentless & Training-free & 24.92 & 12.93 & 15.31 & 11.74 & 63.01 & 19.32 & 27.82 & 18.85\\
  CoSIL & Training-free & \underline{29.30} & 8.98 & 12.90 & 8.07 & \underline{70.12} & 17.90 & 27.39 & 17.42 \\
    LocAgent & Training-free & 17.62 & 11.71 & 12.71 & 10.31 & 60.96 & 34.88 & 40.67 & 33.33\\
  OrcaLoca & Training-free & 27.70 & \underline{20.29} & \underline{21.70} & \underline{17.92} & 48.04 & \underline{48.65} & \underline{47.36} &\underline{45.77}\\
  \midrule
  RepoSearcher & Training-free & 7.86 & 8.08 & 7.48 & 6.85 & 48.61 & 21.15 & 27.36 & 20.39 \\
  RepoSearcher & Distillation+GRPO & \textbf{63.26} & 19.24 & 27.37 & 17.59 & \textbf{84.11} & 19.97 & 31.64 & 19.57 \\
  \midrule
  \rowcolor{yellow!20}
  RepoNavigator & Training-free & 15.89 & 17.46 & 16.19 & 15.46 & 42.36 & 43.23 & 42.12 & 40.97 \\
  % RepoNavigator & RFT & 24.92 & 27.44 & 25.38 & 24.01 & 48.35 & 50.13 & 48.57 & 47.38\\
  \rowcolor{blue!10}
  RepoNavigator & GRPO & 26.69 & \textbf{30.34} & \textbf{27.49} & \textbf{26.43} & 50.62 & \textbf{53.83} & \textbf{51.63} & \textbf{50.62} \\
\midrule
\midrule
& \multicolumn{7}{c}{\textbf{\textit{Qwen2.5-14B}}} \\
\midrule
  Agentless & Training-free & 25.20 & 14.30 & 16.14 & 12.28 & 75.65 & 19.76 & 29.88 & 19.30\\
  CoSIL & Training-free & \textbf{\underline{48.61}} & 13.40 & 19.81 & 12.12 & \textbf{\underline{78.35}} & 18.10 & 28.79 & 17.72  \\
LocAgent & Training-free & 35.62 & 13.32 & 17.71 & 12.32 & 71.42 & 31.66 & 40.77 & 30.64\\

  OrcaLoca & Training-free & 29.92 & 20.98 & 22.77 & 18.92 & 52.17 & 52.15 & 50.93 & 48.72\\
  
  RepoSearcher & Training-free & 26.13 & 11.96 & 14.35 & 10.60 & 74.77 & 18.80 & 28.79 & 18.15\\
  \rowcolor{yellow!20}
  RepoNavigator & Training-free & 27.96 & \underline{25.77} & \underline{25.58} & \underline{23.00} & 59.00 & \underline{56.68} & \underline{56.39} & \underline{53.74}  \\
  % RepoNavigator & RFT &27.37 & \textbf{31.53} & 28.33 & \textbf{27.24} & 57.42 & \textbf{61.61} & 58.58 & \textbf{57.29}\\
  \rowcolor{blue!10}
  RepoNavigator & GRPO & 31.02 & \textbf{30.08} & \textbf{29.23} & \textbf{26.84} & 61.60 & \textbf{58.97} & \textbf{58.90} & \textbf{56.36} \\
\midrule
\midrule
& \multicolumn{7}{c}{\textbf{\textit{Qwen2.5-32B}}} \\
\midrule
Agentless & Training-free & 40.79 & 24.07 & 27.33 & 22.08 & 78.93  & 25.60  &  35.38 &  24.96 \\
  CoSIL & Training-free & 55.38 & 14.85 & 22.11 & 13.52 & 83.50 & 19.34 & 30.77 & 18.93 \\
  
  LocAgent & Training-free & 46.79 & 16.29 & 21.48 & 14.18 & 79.39 & 34.18 & 44.18 &  33.24\\
  OrcaLoca & Training-free & 39.14 & \underline{25.59} & \underline{28.72} & 22.89 & 59.57 & 59.51 & 58.11 & 55.62  \\
  \midrule
  RepoSearcher & Training-free & 18.83 & 19.05 & 18.10 & 16.79 & 66.60 & 39.81 & 45.81 & 37.93  \\
  RepoSearcher & Distillation+GRPO & \textbf{\underline{69.50}} & 20.29 & 29.11 & 18.23 & \textbf{\underline{89.33}} & 20.27 & 32.93 & 20.35  \\
  \midrule
  \rowcolor{yellow!20}
  RepoNavigator & Training-free & 28.11 & 28.19 & 27.12 & \underline{25.16} & 63.05 & \underline{62.75} & \underline{61.67} & \underline{59.28}\\
  \rowcolor{blue!10}
   RepoNavigator & GRPO & 33.71 & \textbf{37.19} & \textbf{34.09} & \textbf{32.30} & 67.29 & \textbf{70.76} & \textbf{67.75} & \textbf{65.75}\\
% \midrule
% \multirow{2}{*}{\textbf{Other Models}} 
%  & LocAgent (Claude3.5) & Training-free & 0.196 & 0.122 & 0.434 & 0.306 \\
%  & RepoSearcher (Claude3.7) & RFT+RL & 0.283 & 0.179 & 0.332 & 0.207 \\
\bottomrule
\end{tabular}
\end{table*}

\section{Experiment}
\label{sec:exp}

\subsection{Experiment Setup}

\paragraph{Datasets}
We extract valid samples from SWE-smith \cite{yang2025swe} to form the training set. We apply Qwen2.5-7B-Instruct with RepoNavigator to sample each data 16 times. A sample is abandoned if all 16 scores are zero. For validation, we test our method on SWE-bench\_Verified \cite{jimenez2023swe}, and the Python subset of SWE-bench\_Pro \cite{yang2025swe}. For ground-truth locations, we directly use the locations in golden patches. All datasets are open-source and are built on real-world GitHub issues. We add an entry point to the user prompt (for RepoNavigator and all baselines) for each issue as the starting context; details are in Appendix~\ref{sec:metrics}.

\paragraph{Metrics}
Previous works \cite{chen-etal-2025-locagent,ma2025toolintegratedreinforcementlearningrepo} applied recall and precision as metrics. However, because the predicted locations and ground-truth locations are sets of strings, recall and precision singularly cannot reflect the performance fairly. Thus, we utilize sample-F1 (which is the averaged score of per-sample F1 values) and IoU (intersection over union) as our core metrics. At the same time, we also present the recall and precision scores to align with previous methods, although they do not reflect the methods' performance fairly.
\begin{table*}[!t]
\centering
\caption{Comparison of different agent pipelines on function-level and file-level metrics on SWE-bench\_Pro. \textbf{Bold numbers} denote the best performance among same-size models; \underline{underline numbers} denote the best training-free performance among same-size models; \colorbox{yellow!20}{yellow background} illustrates training-free RepoNavigator; \colorbox{blue!10}{blue background} illustrates RepoNavigator trained with GRPO.}
\label{tab:agent_comparison_pro}
\small
\setlength{\tabcolsep}{5pt}
\renewcommand{\arraystretch}{1.2}
\begin{tabular}{cc|cccc|cccc}

\toprule
% \multirow{2}{*}{\textbf{Model}} & 
\multirow{2}{*}{\textbf{Agent Pipeline}} & 
\multirow{2}{*}{\textbf{Model}} & 
\multicolumn{4}{c|}{\textbf{Function-level}} &
\multicolumn{4}{c}{\textbf{File-level}} \\
\cmidrule(lr){3-6} \cmidrule(lr){7-10}
 & & {Recall} & {Precision} & {Sample-F1} & {IoU} 
& {Recall} & {Precision} & {Sample-F1} & {IoU} \\

\midrule
\midrule
& \multicolumn{7}{c}{\textbf{\textit{Qwen2.5-7B}}} \\
\midrule
    Agentless & Training-free & 12.82 & 6.94 & 8.05 & 5.73 & \textbf{\underline{39.41}} & 13.15 & 18.89 & 12.35\\
  CoSIL & Training-free & 8.64 & 3.33 & 4.58 & 2.87 & 26.64 & 8.47 & 12.11 & 7.70 \\
LocAgent & Training-free & 1.01 & 0.02 & 0.65 & 0.40 & 12.16 & 0.17 & 10.81 & 8.93 \\
  LocAgent & Distillation & \textbf{\underline{13.76}} & 1.61 & 2.66 & 1.48 & 37.44 & 5.11 & 8.39 & 5.00 \\
  RepoSearcher & Training-free & 1.07 & 0.93 & 0.97 & 0.86 & 4.91 & 1.64 & 2.30 & 1.63 \\
  \rowcolor{yellow!20}
  RepoNavigator & Training-free & 9.84 & \underline{14.65} & \underline{10.67} & \underline{9.20} & 30.50 & \underline{37.24} & \underline{31.86} & \underline{28.82} \\
  \rowcolor{blue!10}
  RepoNavigator & GRPO & 12.33 & \textbf{21.26} & \textbf{14.29} & \textbf{12.02} & 36.36 & \textbf{48.13} & \textbf{39.74} & \textbf{36.36} \\
\midrule
\midrule
& \multicolumn{7}{c}{\textbf{\textit{Qwen2.5-14B}}} \\
\midrule
  Agentless & Training-free & 10.49 & 6.75 & 7.41 & 5.28 & 41.42 & 13.42 & 19.02 & 12.37\\
  CoSIL & Training-free & 10.73 & 4.67 & 5.96 & 3.94 & 34.31 & 9.97 & 14.81 & 9.30  \\
LocAgent & Training-free & 6.22 & 0.13 & 3.65 & 2.65 & 15.58 & 0.21 & 11.69 & 9.53 \\
  
  RepoSearcher & Training-free & 2.79 & 1.38 & 1.69 & 1.14 & 17.37 & 5.17 & 7.60 & 4.84\\
  \rowcolor{yellow!20}
  RepoNavigator & Training-free & \underline{14.36} & \underline{19.74} & \underline{15.27} & \underline{12.00} & \underline{43.57} & \underline{54.52} & \underline{46.06} & \underline{41.07}  \\
  \rowcolor{blue!10}
  RepoNavigator & GRPO & \textbf{16.05} & \textbf{25.25} & \textbf{18.06} & \textbf{14.58} & \textbf{46.85} & \textbf{58.64} & \textbf{49.72} & \textbf{45.14} \\
\midrule
\midrule
& \multicolumn{7}{c}{\textbf{\textit{Qwen2.5-32B}}} \\
\midrule
    Agentless & Training-free & 11.08 & 7.31 & 7.98 & 5.80 & 43.07  & 13.89  &  20.07 &  13.11 \\
  CoSIL & Training-free & \underline{14.03} & 6.00 & 7.67 & 4.92 & 48.87 & 14.03 & 20.95 & 13.27 \\
LocAgent & Training-free & 8.72 & 0.17 & 4.30 & 2.90 & 25.73 & 0.38 & 19.77 & 16.50 \\
  RepoSearcher & Training-free & 2.52 & 2.46 & 2.31 & 1.75 & 9.00 & 2.52 & 3.81 & 2.41  \\
  \rowcolor{yellow!20}
  RepoNavigator & Training-free & 13.96 & \underline{20.25} & \underline{15.36} & \underline{12.87} & \underline{50.24} & \underline{63.24} & \underline{53.48} & \underline{48.50}\\
  \rowcolor{blue!10}
   RepoNavigator & GRPO & \textbf{18.13} & \textbf{29.44} & \textbf{20.72} & \textbf{17.16} &\textbf{ 53.49} & \textbf{68.69} & \textbf{57.57} & \textbf{52.44}\\
% \midrule
% \multirow{2}{*}{\textbf{Other Models}} 
%  & LocAgent (Claude3.5) & Training-free & 0.196 & 0.122 & 0.434 & 0.306 \\
%  & RepoSearcher (Claude3.7) & RFT+RL & 0.283 & 0.179 & 0.332 & 0.207 \\
\bottomrule
\end{tabular}
\end{table*}
\paragraph{Training}
For the 7B model, we conduct GRPO with 8 NVIDIA Tesla-A100-80G GPUs. For the 14B and 32B models, we train them with 16 NVIDIA Tesla-A100-80G GPUs. We apply verl \cite{shen2024llm} as the training framework, and we apply vLLM \cite{kwon2023efficient} as the inference engine. We train the model for 1 epoch, while the training batch size is fixed to 128 on 4k training samples filtered from SWE-smith, with maximum prompt length and max response length both set to 10240. Additionally, the group size \(G\) is set to 8, and the temperature is set to 1.0 to encourage exploration. We use greedy decoding in the inference stage to ensure stable performance. More experimental details are provided in Appendix~\ref{sec:more_details}.

\subsection{Effectiveness}
\paragraph{Baselines}
We compare our method against LocAgent \cite{chen-etal-2025-locagent}, CoSIL \cite{jiang2025cosil}, Agentless \cite{xia2024agentless}, OrcaLoca \cite{yu2025orcalocallmagentframework}, and RepoSearcher \cite{ma2025toolintegratedreinforcementlearningrepo}. LocAgent is distilled from Claude3.5-Sonnet, and RepoSearcher is distilled from Claude3.7-Sonnet. See Appendix~\ref{sec:baselines} for more details.
\paragraph{Results}
As illustrated in Table~\ref{tab:agent_comparison_verified}, on balanced metrics (Sample-F1 and IoU) for both function-level and file-level localization, our method surpasses all baseline methods with the same model size. Moreover, if we train RepoNavigator with GRPO, the 7B model surpasses 14B baselines, and our 14B model surpasses 32B baselines on Sample-F1 and IoU. This further validates the effectiveness of RepoNavigator. For 14B and 32B models, RepoNavigator achieves SOTA among all training-free methods. This implies that the tool we implement is effective in practice, and our single tool pipeline is better than previous multi-tool pipelines.

Compared to RepoSearcher, which is distilled from claude-3.7-sonnet \cite{anthropic2025claude37sonnet} and reinforced by GRPO, trained RepoNavigator outperforms it in all metrics except recall. Moreover, we found that our training-free method outperforms RepoSearcher for 14B models. We attribute this improvement to the simplified tool we integrate into the agent (see Sec.~\ref{sec:dicussion} for more details).

To assess the generalizability of RepoNavigator, we present its performance on Python samples from the SWE-bench\_Pro dataset \cite{yang2025swe} in Table~\ref{tab:agent_comparison_pro}. The results on this dataset are consistent with those observed on SWE-bench\_Verified. While we cannot fully exclude the potential influence of data leakage in SWE-bench\_Verified, we can make a stronger claim regarding SWE-bench\_Pro, as it was released after the publication of the Qwen2.5 series.
\begin{figure}[!t]

    \centering
    
    \includegraphics[width=\linewidth]{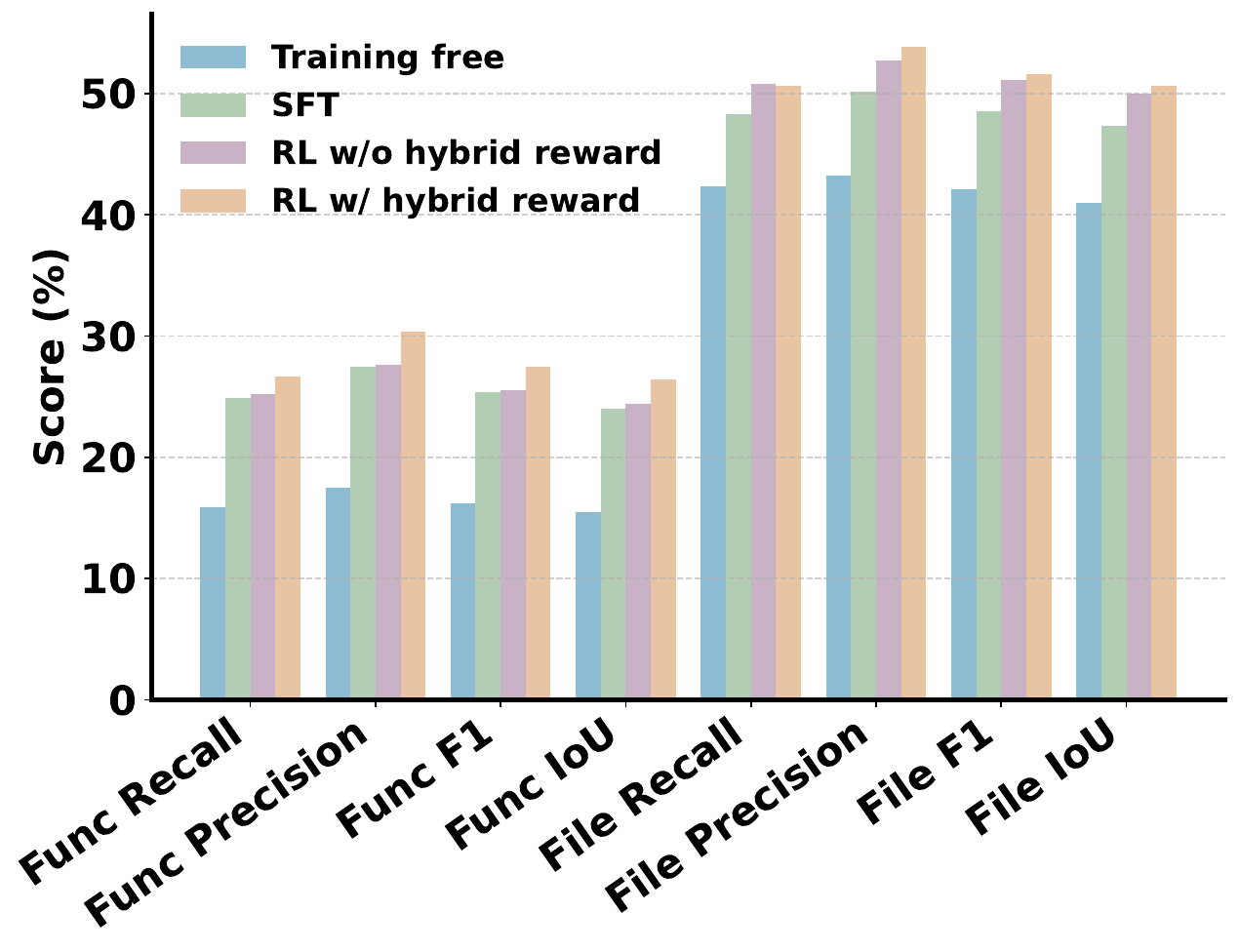}
    \caption{Training strategies comparison: comparison between RepoNavigator with training-free, RFT, GRPO with pure outcome and hybrid reward on Qwen2.5-7B-Instruct.}
    \label{fig:ablation_study}
\end{figure}
\begin{figure}[!t]

    % \vspace{-4pt}  % 减少顶部空白
    \centering
    \includegraphics[width=\linewidth]{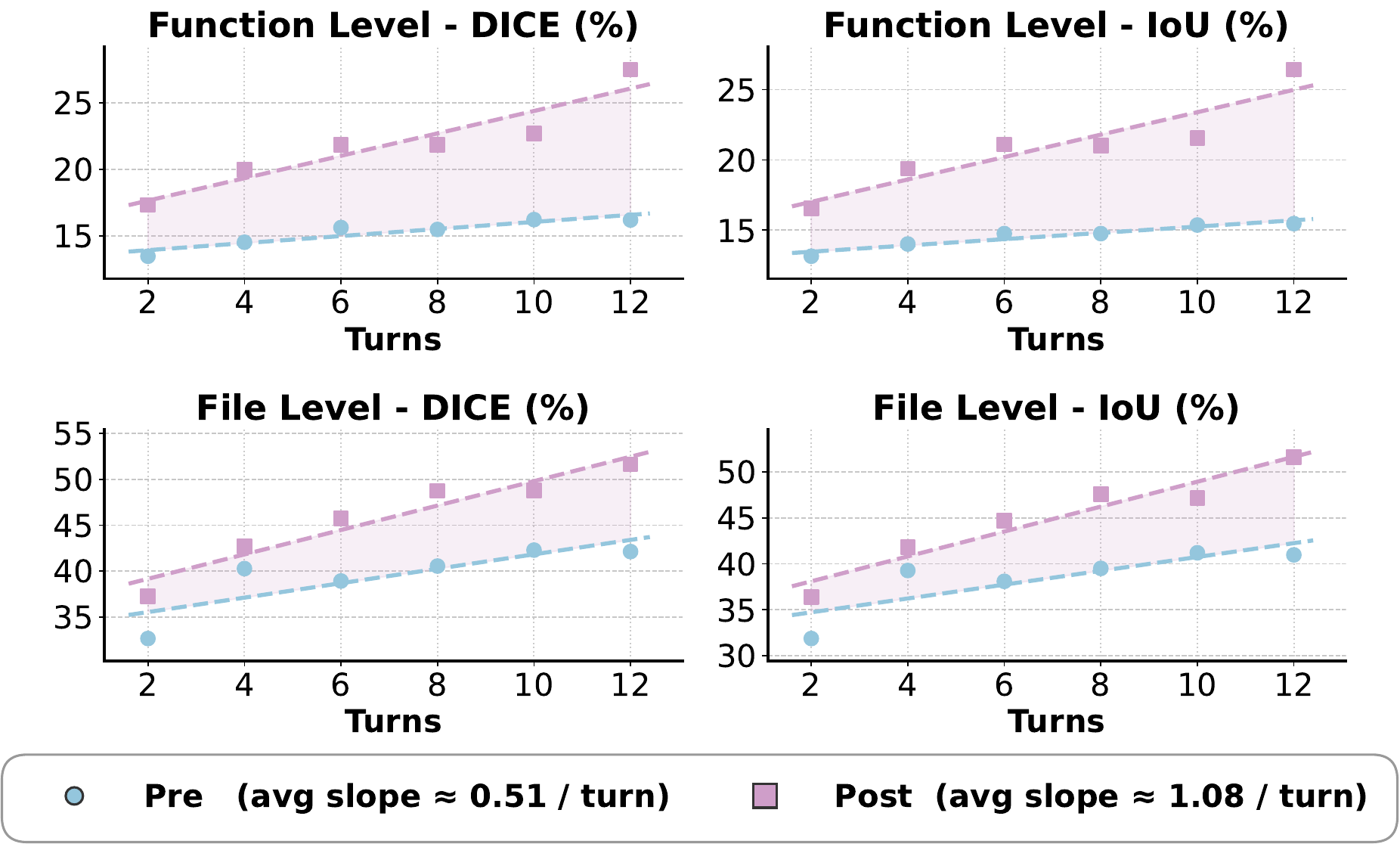}
    \caption{Scaling law of tool-calling, where \textit{Pre} and \textit{Post} denote the corresponding metric before and after training.}
    \label{fig:scaling_law}
\end{figure}

\subsection{Training Strategy Comparison}
We investigate the capabilities of GRPO and our proposed hybrid reward function within the context of agentic training, using SFT as a primary benchmark. Experimental results presented in Fig.~\ref{fig:ablation_study} indicate that our method significantly surpasses both SFT and GRPO with a vanilla reward function, validating the critical role of RL and hybrid reward mechanisms in enhancing agentic capabilities. While \cite{ma2025toolintegratedreinforcementlearningrepo} emphasizes the necessity of an initial SFT phase, our ablation study (detailed in Appendix \ref{sec:ablation_study}) reveals the potential of training fully autonomous agents directly through RL. This observation challenges the requirement for a ``cold start" and resonates with a growing body of literature advocating for RL-centric training paradigms \cite{liu2024deepseek,kirk2023understanding,jin2025rl}.

We also remove the success rate in the reward function for ablation. As presented in Fig.~\ref{fig:ablation_study}, reinforcement learning with a hybrid reward (with tool-calling success rate) has higher performance than pure outcome reward (without tool-calling success rate). This indicates that learning to correctly call tools is vital in agentic learning.
\subsection{Scaling Law of Tool-Calling}

To assess the significance of tool-calling in RepoNavigator, we varied the maximum number of tool-calling turns and reported the results in Fig.~\ref{fig:scaling_law}. As shown in the figure, allowing more tool-calling turns generally leads to improved performance for RepoNavigator, both before and after reinforcement learning (RL) training. These results empirically validate the scaling law of tool-calling in this context.
\begin{table}[!t]
\small
    \centering
        \caption{Downstream influence: We use Qwen2.5-14B-Instruct on SWE-bench\_Verified to evaluate issue resolution.}
    % \resizebox{\linewidth}{!}{%
        \begin{tabular}{ccccc}
        \toprule
            \textbf{Agent Pipeline} & \textbf{Func-IoU(\%)} & \textbf{Resolved(\%)} \\
            \midrule
            Agentless & 12.28 & 10.12\\
            LocAgent & 12.32 &  13.01\\
            RepoSearcher & 10.60 &  12.47\\
            RepoNavigator & 23.00 & 14.74\\
            \rowcolor{blue!10}
            RepoNavigator+RL & 26.84 &  15.03\\
        \bottomrule
        \end{tabular}
    % }% end resizebox   

\label{tab:resolve}

\end{table}
\begin{figure}[!t]
    \centering
    \includegraphics[width=0.999\linewidth]{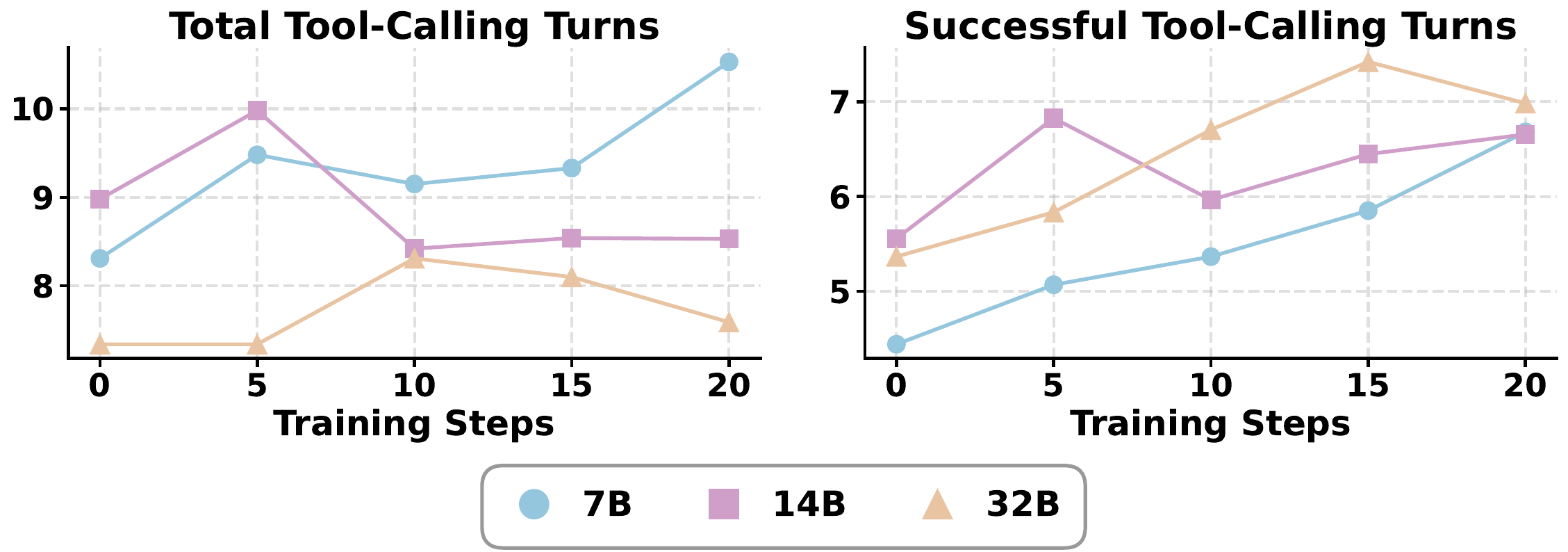}
    \caption{Training dynamics of tool-calling behaviors.}
    \label{fig:turns}
\end{figure}
\subsection{Evolution of tool-calling proficiency}
A potential side effect of incorporating success rates into the GRPO reward function is the emergence of conservative agents that artificially inflate success metrics by suppressing tool usage. However, the empirical results in Fig.~\ref{fig:turns} alleviate this concern. Specifically, the number of successful tool-calling turns increases, while the total number of turns remains stable or even rises. For the tool success rate, Fig.~\ref{fig:tool_reward} in the Appendix illustrates a rising trend. Notably, the 7B model exhibits a pronounced increase in total turns. In contrast to the 14B and 32B models—which leverage stronger reasoning capabilities to reach targets through fewer but more precise tool invocations—the 7B model undergoes a more extensive exploration phase. This behavior can be attributed to its initially weaker tool mastery, as evidenced by the lower tool success rates shown in Fig.~\ref{fig:tool_reward} in the Appendix.
\subsection{Influence on Issue Resolution}
To assess how different localization strategies affect downstream issue resolution, we evaluate RepoNavigator against baseline methods on SWE-bench\_Verified. Specifically, we adopt the repair phase of Agentless and substitute its localization module with alternative approaches. The results are summarized in Table~\ref{tab:resolve}.
RepoNavigator consistently outperforms all baselines in terms of issue resolution accuracy, and the incorporation of reinforcement learning further enhances its performance.

\section{Discussion: Building Fewer yet More Capable Tools}
\label{sec:dicussion}
In this section, we analyze the logic behind RepoNavigator: A minimalist, unified toolset is more effective than a multitude of fragmented, narrowly scoped tools.

\subsection{Impact on the Action Space of Agents}

Let the total number of available tools be denoted as $k$. When only a single tool—specifically the \texttt{jump} tool—is retained, the system’s structural relations become simpler, as both the action space and the observation space are restricted to what this tool can access. In this case, the set of possible actions and observable elements is smaller than when multiple tools are available. This reduction is generally beneficial, since additional tools often introduce new and unfamiliar interfaces that large language models have not been exposed to during pretraining.

\subsection{Impact on Tool-Calling Success Rate}

For a given process in issue localization (for instance, checking the code snippet of a function), let the success probability of the $i$-th call be $p_i$.
For a task that requires $k$ sequential tool calls, the overall success rate can be expressed as:
\begin{equation}
P_\text{succ}(k) = \prod_{i=1}^{k} p_i.
\end{equation}
The cumulative success rate typically decreases as the number of required tool calls increases.
Therefore, in general, completing a task with a single, more versatile tool tends to be more reliable than relying on multiple narrow-scope tools executed in sequence.

\subsection{Impact on the Prediction Space}
\begin{figure}
    \centering
    \includegraphics[width=0.9\linewidth]{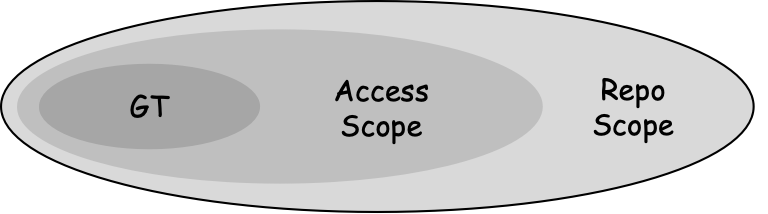}
    \caption{Venn graph illustrating the access scope of \texttt{jump}. Compared with the repository scope, the access scope of \texttt{jump} has a much higher IoU with the groundtruth set.}
    \label{fig:tas}
\end{figure}
The access scope of a tool is defined as the complete set of files, symbols, and other resources that the tool can access within a repository. For a \texttt{jump} tool that navigates to symbol definitions, its access scope can be obtained by starting from a given entry point and recursively resolving all referenced symbols until no new definitions can be reached. Apparently, its access scope is significantly smaller than the full repository scope. Consequently, when computing the Intersection over Union (IoU) between the prediction set and the groundtruth set, using the \texttt{jump} tool results in a higher IoU, as depicted in Fig.~\ref{fig:tas}. On the other hand, applying multiple repo-level retrieval tools results in the access scope equal to the whole repository scope. 

Starting from a designated entry point, the recursive application of the \texttt{jump} operation—which resolves the definitions of all referenced symbols—effectively constructs a transitive closure of all semantically activated symbols. Since any code location contributing to a specific issue must inherently reside on a dependency path originating from the entry point, it is logically reachable through this recursive expansion of symbol references. Consequently, the final access scope generated by an exhaustive \texttt{jump} traversal is guaranteed to encompass all locations requiring modification to resolve the issue, excluding only those introduced by dynamic, run-time changes to the repository (e.g., monkey-patching).

\subsection{Verification}
To further validate this hypothesis, we modify the tool set of RepoNavigator and perform reinforcement learning with only outcome-based rewards. We add additional tools commonly used in prior work~\cite{chen-etal-2025-locagent, ma2025toolintegratedreinforcementlearningrepo, jiang2025cosil}. As shown in Table~\ref{tab:m_vs_s} (full results in Table~\ref{tab:m_vs_s_full}), additional tools hinders performance. This suggests that \textbf{fewer yet more capable tools} are preferable.
\begin{table}
\small
    \centering
        \caption{We change the tool set of RepoNavigator and present the function-level IoU (\%) on Qwen2.5-7B-Instruct. From the table, we demonstrate that excessive tools do not boost RepoNavigator's performance.}
    %\resizebox{\linewidth}{!}{%
        \begin{tabular}{cccc|c}
        \toprule
            \textbf{Jump} & \textbf{GetClass} & \textbf{GetFunc} & \textbf{GetStruc} & \textbf{IoU}\\
            \midrule
            \cmark & \cmark & \cmark & \cmark & 13.71\\
            
            \cmark & \cmark & \cmark & \xmark & 21.44\\
            \cmark & \xmark & \xmark & \cmark & 24.00\\
            \rowcolor{blue!10}
            \cmark & \xmark & \xmark & \xmark & 24.28\\
        \bottomrule
        \end{tabular}
    \label{tab:m_vs_s}
\end{table}

\section{Conclusion} % 这一段不需要有引用吗？
In this work, we propose RepoNavigator, a repository-level issue localization agent that replaces complex multi-tool pipelines with a single, more capable \texttt{jump} tool for symbol resolution. This unified design better reflects real code execution flow while reducing the brittleness of tool chaining. Trained with tool-integrated GRPO, RepoNavigator jointly learns reasoning and tool invocation in a closed loop, enabling end-to-end optimization without closed-source teacher models or distillation.

Extensive experiments across SWE-bench\_Verified and SWE-bench\_Pro demonstrate that RepoNavigator achieves state-of-the-art localization performance.  We theoretically analyze the results, confirming that a single powerful tool, jointly optimized with reinforcement learning, can provide stronger robustness and more reliable multi-step reasoning than previous frameworks relying on multiple narrowly scoped tools. Future work will explore extending RepoNavigator from Python to more programming languages.
\section*{Acknowledgement}
This work is supported by the Zhongguancun Academy (Grant No.s C20250210). We thank Chenming Tang (Peking University) for the constructive suggestions provided during the discussion with the authors.
\section*{Impact Statement}
This paper presents RepoNavigator, an LLM agent designed to advance repository-level issue localization by introducing a streamlined, single-tool navigation paradigm and an autonomous Reinforcement Learning training framework that enhances the efficiency and accessibility of automated software maintenance. While this technology significantly reduces manual debugging overhead and democratizes complex codebase navigation, its precision in identifying functional weak points could potentially be repurposed for malicious intent, such as accelerating the discovery of software vulnerabilities for targeted cyberattacks. Furthermore, its integration into enterprise environments necessitates rigorous compliance frameworks to ensure that all AI-driven code modifications are meticulously tracked, auditable, and aligned with security standards to mitigate technical debt or legal liabilities.
\nocite{langley00}
% \newpage
\bibliography{example_paper}

@inproceedings{langley00,
 author    = {P. Langley},
 title     = {Crafting Papers on Machine Learning},
 year      = {2000},
 pages     = {1207--1216},
 editor    = {Pat Langley},
 booktitle     = {Proceedings of the 17th International Conference
              on Machine Learning (ICML 2000)},
 address   = {Stanford, CA},
 publisher = {Morgan Kaufmann}
}

@article{liu2024deepseek,
  title={Deepseek-v3 technical report},
  author={Liu, Aixin and Feng, Bei and Xue, Bing and Wang, Bingxuan and Wu, Bochao and Lu, Chengda and Zhao, Chenggang and Deng, Chengqi and Zhang, Chenyu and Ruan, Chong and others},
  journal={arXiv preprint arXiv:2412.19437},
  year={2024}
}

@article{yang2025qwen3,
  title={Qwen3 technical report},
  author={Yang, An and Li, Anfeng and Yang, Baosong and Zhang, Beichen and Hui, Binyuan and Zheng, Bo and Yu, Bowen and Gao, Chang and Huang, Chengen and Lv, Chenxu and others},
  journal={arXiv preprint arXiv:2505.09388},
  year={2025}
}

@article{team2024qwen2,
  title={Qwen2 technical report},
  author={Team, Qwen},
  journal={arXiv preprint arXiv:2407.10671},
  year={2024}
}

@article{huang2024understanding,
  title={Understanding the planning of LLM agents: A survey},
  author={Huang, Xu and Liu, Weiwen and Chen, Xiaolong and Wang, Xingmei and Wang, Hao and Lian, Defu and Wang, Yasheng and Tang, Ruiming and Chen, Enhong},
  journal={arXiv preprint arXiv:2402.02716},
  year={2024}
}

@article{guo2024large,
  title={Large language model based multi-agents: A survey of progress and challenges},
  author={Guo, Taicheng and Chen, Xiuying and Wang, Yaqi and Chang, Ruidi and Pei, Shichao and Chawla, Nitesh V and Wiest, Olaf and Zhang, Xiangliang},
  journal={arXiv preprint arXiv:2402.01680},
  year={2024}
}

@article{jimenez2023swe,
  title={Swe-bench: Can language models resolve real-world github issues?},
  author={Jimenez, Carlos E and Yang, John and Wettig, Alexander and Yao, Shunyu and Pei, Kexin and Press, Ofir and Narasimhan, Karthik},
  journal={arXiv preprint arXiv:2310.06770},
  year={2023}
}

@inproceedings{yang2024sweagent,
  title={{SWE}-agent: Agent-Computer Interfaces Enable Automated Software Engineering},
  author={John Yang and Carlos E Jimenez and Alexander Wettig and Kilian Lieret and Shunyu Yao and Karthik R Narasimhan and Ofir Press},
  booktitle={The Thirty-eighth Annual Conference on Neural Information Processing Systems},
  year={2024},
  url={https://arxiv.org/abs/2405.15793}
}

@inproceedings{
  wang2025openhands,
  title={OpenHands: An Open Platform for {AI} Software Developers as Generalist Agents},
  author={Xingyao Wang and Boxuan Li and Yufan Song and Frank F. Xu and Xiangru Tang and Mingchen Zhuge and Jiayi Pan and Yueqi Song and Bowen Li and Jaskirat Singh and Hoang H. Tran and Fuqiang Li and Ren Ma and Mingzhang Zheng and Bill Qian and Yanjun Shao and Niklas Muennighoff and Yizhe Zhang and Binyuan Hui and Junyang Lin and Robert Brennan and Hao Peng and Heng Ji and Graham Neubig},
  booktitle={The Thirteenth International Conference on Learning Representations},
  year={2025},
  url={https://openreview.net/forum?id=OJd3ayDDoF}
}

@misc{SWESwiss2025,
    title={SWE-Swiss: A Multi-Task Fine-Tuning and RL Recipe for High-Performance Issue Resolution},
    author={He, Zhenyu and Yang, Qingping and Sheng, Wei and Zhong, Xiaojian and Zhang, Kechi and An, Chenxin and Shi, Wenlei and Cai, Tianle and He, Di and Chen, Jiaze and Xu, Jingjing},
    howpublished={https://github.com/zhenyuhe00/SWE-Swiss},
    note={Notion Blog},
    year={2025}
}

@misc{ma2025toolintegratedreinforcementlearningrepo,
      title={Tool-integrated Reinforcement Learning for Repo Deep Search}, 
      author={Zexiong Ma and Chao Peng and Qunhong Zeng and Pengfei Gao and Yanzhen Zou and Bing Xie},
      year={2025},
      eprint={2508.03012},
      archivePrefix={arXiv},
      primaryClass={cs.SE},
      url={https://arxiv.org/abs/2508.03012}, 
}

@inproceedings{chen-etal-2025-locagent,
 title = "{L}oc{A}gent: Graph-Guided {LLM} Agents for Code Localization",
 author = "Chen, Zhaoling  and
   Tang, Robert  and
   Deng, Gangda  and
   Wu, Fang  and
   Wu, Jialong  and
   Jiang, Zhiwei  and
   Prasanna, Viktor  and
   Cohan, Arman  and
   Wang, Xingyao",
 editor = "Che, Wanxiang  and
   Nabende, Joyce  and
   Shutova, Ekaterina  and
   Pilehvar, Mohammad Taher",
 booktitle = "Proceedings of the 63rd Annual Meeting of the Association for Computational Linguistics (Volume 1: Long Papers)",
 month = jul,
 year = "2025",
 address = "Vienna, Austria",
 publisher = "Association for Computational Linguistics",
 url = "https://aclanthology.org/2025.acl-long.426/",
 doi = "10.18653/v1/2025.acl-long.426",
 pages = "8697--8727",
 ISBN = "979-8-89176-251-0"
}

@misc{yu2025orcalocallmagentframework,
      title={OrcaLoca: An LLM Agent Framework for Software Issue Localization},
      author={Zhongming Yu and Hejia Zhang and Yujie Zhao and Hanxian Huang and Matrix Yao and Ke Ding and Jishen Zhao},
      year={2025},
      eprint={2502.00350},
      archivePrefix={arXiv},
      primaryClass={cs.SE},
      url={https://arxiv.org/abs/2502.00350},
}

@article{shen2024llm,
  title={Llm with tools: A survey},
  author={Shen, Zhuocheng},
  journal={arXiv preprint arXiv:2409.18807},
  year={2024}
}

@article{yuan2024easytool,
  title={Easytool: Enhancing llm-based agents with concise tool instruction},
  author={Yuan, Siyu and Song, Kaitao and Chen, Jiangjie and Tan, Xu and Shen, Yongliang and Kan, Ren and Li, Dongsheng and Yang, Deqing},
  journal={arXiv preprint arXiv:2401.06201},
  year={2024}
}

@article{lu2024toolsandbox,
  title={Toolsandbox: A stateful, conversational, interactive evaluation benchmark for llm tool use capabilities},
  author={Lu, Jiarui and Holleis, Thomas and Zhang, Yizhe and Aumayer, Bernhard and Nan, Feng and Bai, Felix and Ma, Shuang and Ma, Shen and Li, Mengyu and Yin, Guoli and others},
  journal={arXiv preprint arXiv:2408.04682},
  year={2024}
}

@article{hui2024qwen2,
  title={Qwen2. 5-coder technical report},
  author={Hui, Binyuan and Yang, Jian and Cui, Zeyu and Yang, Jiaxi and Liu, Dayiheng and Zhang, Lei and Liu, Tianyu and Zhang, Jiajun and Yu, Bowen and Lu, Keming and others},
  journal={arXiv preprint arXiv:2409.12186},
  year={2024}
}

@article{guo2024deepseek,
  title={DeepSeek-Coder: When the Large Language Model Meets Programming--The Rise of Code Intelligence},
  author={Guo, Daya and Zhu, Qihao and Yang, Dejian and Xie, Zhenda and Dong, Kai and Zhang, Wentao and Chen, Guanting and Bi, Xiao and Wu, Yu and Li, YK and others},
  journal={arXiv preprint arXiv:2401.14196},
  year={2024}
}

@inproceedings{liu2024dynamic,
  title={A dynamic LLM-powered agent network for task-oriented agent collaboration},
  author={Liu, Zijun and Zhang, Yanzhe and Li, Peng and Liu, Yang and Yang, Diyi},
  booktitle={First Conference on Language Modeling},
  year={2024}
}

@article{schmidgall2025agent,
  title={Agent laboratory: Using llm agents as research assistants},
  author={Schmidgall, Samuel and Su, Yusheng and Wang, Ze and Sun, Ximeng and Wu, Jialian and Yu, Xiaodong and Liu, Jiang and Moor, Michael and Liu, Zicheng and Barsoum, Emad},
  journal={arXiv preprint arXiv:2501.04227},
  year={2025}
}

@article{jin2025search,
  title={Search-r1: Training llms to reason and leverage search engines with reinforcement learning},
  author={Jin, Bowen and Zeng, Hansi and Yue, Zhenrui and Yoon, Jinsung and Arik, Sercan and Wang, Dong and Zamani, Hamed and Han, Jiawei},
  journal={arXiv preprint arXiv:2503.09516},
  year={2025}
}

@inproceedings{hong2024cogagent,
  title={Cogagent: A visual language model for gui agents},
  author={Hong, Wenyi and Wang, Weihan and Lv, Qingsong and Xu, Jiazheng and Yu, Wenmeng and Ji, Junhui and Wang, Yan and Wang, Zihan and Dong, Yuxiao and Ding, Ming and others},
  booktitle={Proceedings of the IEEE/CVF Conference on Computer Vision and Pattern Recognition},
  pages={14281--14290},
  year={2024}
}

@article{yan2025mathagent,
  title={Mathagent: Leveraging a mixture-of-math-agent framework for real-world multimodal mathematical error detection},
  author={Yan, Yibo and Wang, Shen and Huo, Jiahao and Yu, Philip S and Hu, Xuming and Wen, Qingsong},
  journal={arXiv preprint arXiv:2503.18132},
  year={2025}
}

@article{yang2024swe,
  title={Swe-bench multimodal: Do ai systems generalize to visual software domains?},
  author={Yang, John and Jimenez, Carlos E and Zhang, Alex L and Lieret, Kilian and Yang, Joyce and Wu, Xindi and Press, Ori and Muennighoff, Niklas and Synnaeve, Gabriel and Narasimhan, Karthik R and others},
  journal={arXiv preprint arXiv:2410.03859},
  year={2024}
}

@article{xia2024agentless,
  title={Agentless: Demystifying llm-based software engineering agents},
  author={Xia, Chunqiu Steven and Deng, Yinlin and Dunn, Soren and Zhang, Lingming},
  journal={arXiv preprint arXiv:2407.01489},
  year={2024}
}

@article{li2024personal,
  title={Personal llm agents: Insights and survey about the capability, efficiency and security},
  author={Li, Yuanchun and Wen, Hao and Wang, Weijun and Li, Xiangyu and Yuan, Yizhen and Liu, Guohong and Liu, Jiacheng and Xu, Wenxing and Wang, Xiang and Sun, Yi and others},
  journal={arXiv preprint arXiv:2401.05459},
  year={2024}
}

@article{ahn2024large,
  title={Large language models for mathematical reasoning: Progresses and challenges},
  author={Ahn, Janice and Verma, Rishu and Lou, Renze and Liu, Di and Zhang, Rui and Yin, Wenpeng},
  journal={arXiv preprint arXiv:2402.00157},
  year={2024}
}

@inproceedings{gupta2023visual,
  title={Visual programming: Compositional visual reasoning without training},
  author={Gupta, Tanmay and Kembhavi, Aniruddha},
  booktitle={Proceedings of the IEEE/CVF conference on computer vision and pattern recognition},
  pages={14953--14962},
  year={2023}
}

@article{yang2025swe,
  title={Swe-smith: Scaling data for software engineering agents},
  author={Yang, John and Lieret, Kilian and Jimenez, Carlos E and Wettig, Alexander and Khandpur, Kabir and Zhang, Yanzhe and Hui, Binyuan and Press, Ofir and Schmidt, Ludwig and Yang, Diyi},
  journal={arXiv preprint arXiv:2504.21798},
  year={2025}
}

@article{jiang2025cosil,
  title={CoSIL: Software Issue Localization via LLM-Driven Code Repository Graph Searching},
  author={Jiang, Zhonghao and Ren, Xiaoxue and Yan, Meng and Jiang, Wei and Li, Yong and Liu, Zhongxin},
  journal={arXiv preprint arXiv:2503.22424},
  year={2025}
}

@inproceedings{kwon2023efficient,
  title={Efficient Memory Management for Large Language Model Serving with PagedAttention},
  author={Woosuk Kwon and Zhuohan Li and Siyuan Zhuang and Ying Sheng and Lianmin Zheng and Cody Hao Yu and Joseph E. Gonzalez and Hao Zhang and Ion Stoica},
  booktitle={Proceedings of the ACM SIGOPS 29th Symposium on Operating Systems Principles},
  year={2023}
}

@misc{anthropic2025claude37sonnet,
  author       = {Anthropic},
  title        = {Claude 3.7 Sonnet and Claude Code},
  howpublished = {\url{https://www.anthropic.com/news/claude-3-7-sonnet}},
  year         = {2025},
  month        = feb,
  day          = {24},
  note         = {date: 2025-11-18}
}

@article{wang2025extracting,
  title={Extracting Conceptual Knowledge to Locate Software Issues},
  author={Wang, Ying and Mao, Wenjun and Wang, Chong and Zhou, Zhenhao and Zhou, Yicheng and Zhao, Wenyun and Lou, Yiling and Peng, Xin},
  journal={arXiv preprint arXiv:2509.21427},
  year={2025}
}

@article{yu2025dapo,
  title={Dapo: An open-source llm reinforcement learning system at scale},
  author={Yu, Qiying and Zhang, Zheng and Zhu, Ruofei and Yuan, Yufeng and Zuo, Xiaochen and Yue, Yu and Dai, Weinan and Fan, Tiantian and Liu, Gaohong and Liu, Lingjun and others},
  journal={arXiv preprint arXiv:2503.14476},
  year={2025}
}

@misc{deepswe2025,
title={DeepSWE: Training a State-of-the-Art Coding Agent from Scratch by Scaling RL},
author={Michael Luo and Naman Jain and Jaskirat Singh and Sijun Tan and Ameen Patel and Qingyang Wu and Alpay Ariyak and Colin Cai and Tarun Venkat, Shang Zhu and Ben Athiwaratkun and Manan Roongta and Ce Zhang and Li Erran Li and Raluca Ada Popa and Koushik Sen and Ion Stoica},
url={https://pretty-radio-b75.notion.site/DeepSWE-Training-a-Fully-Open-sourced-State-of-the-Art-Coding-Agent-by-Scaling-RL-22281902c1468193aabbe9a8c59bbe33},
    note={Notion Blog},
year={2025}
}

@article{yue2025vapo,
  title={Vapo: Efficient and reliable reinforcement learning for advanced reasoning tasks},
  author={Yue, Yu and Yuan, Yufeng and Yu, Qiying and Zuo, Xiaochen and Zhu, Ruofei and Xu, Wenyuan and Chen, Jiaze and Wang, Chengyi and Fan, TianTian and Du, Zhengyin and others},
  journal={arXiv preprint arXiv:2504.05118},
  year={2025}
}

@inproceedings{sohrabizadehnemotron,
  title={Nemotron-CORTEXA: Enhancing LLM Agents for Software Engineering Tasks via Improved Localization and Solution Diversity},
  author={Sohrabizadeh, Atefeh and Song, Jialin and Liu, Mingjie and Roy, Rajarshi and Lee, Chankyu and Raiman, Jonathan and Catanzaro, Bryan},
  booktitle={Forty-second International Conference on Machine Learning}
}

@article{pan2024training,
  title={Training software engineering agents and verifiers with swe-gym},
  author={Pan, Jiayi and Wang, Xingyao and Neubig, Graham and Jaitly, Navdeep and Ji, Heng and Suhr, Alane and Zhang, Yizhe},
  journal={arXiv preprint arXiv:2412.21139},
  year={2024}
}

@article{li2025patchpilot,
  title={PatchPilot: A Cost-Efficient Software Engineering Agent with Early Attempts on Formal Verification},
  author={Li, Hongwei and Tang, Yuheng and Wang, Shiqi and Guo, Wenbo},
  journal={arXiv preprint arXiv:2502.02747},
  year={2025}
}

@inproceedings{liu2025codexgraph,
  title={Codexgraph: Bridging large language models and code repositories via code graph databases},
  author={Liu, Xiangyan and Lan, Bo and Hu, Zhiyuan and Liu, Yang and Zhang, Zhicheng and Wang, Fei and Shieh, Michael Qizhe and Zhou, Wenmeng},
  booktitle={Proceedings of the 2025 Conference of the Nations of the Americas Chapter of the Association for Computational Linguistics: Human Language Technologies (Volume 1: Long Papers)},
  pages={142--160},
  year={2025}
}

@inproceedings{xiang2025promptsculptor,
  title={Promptsculptor: Multi-agent based text-to-image prompt optimization},
  author={Xiang, Dawei and Xu, Wenyan and Chu, Kexin and Ding, Tianqi and Shen, Zixu and Zeng, Yiming and Su, Jianchang and Zhang, Wei},
  booktitle={Proceedings of the 2025 Conference on Empirical Methods in Natural Language Processing: System Demonstrations},
  pages={774--786},
  year={2025}
}

@article{wang2023promptagent,
  title={Promptagent: Strategic planning with language models enables expert-level prompt optimization},
  author={Wang, Xinyuan and Li, Chenxi and Wang, Zhen and Bai, Fan and Luo, Haotian and Zhang, Jiayou and Jojic, Nebojsa and Xing, Eric P and Hu, Zhiting},
  journal={arXiv preprint arXiv:2310.16427},
  year={2023}
}

@inproceedings{chen2024comm,
  title={Comm: Collaborative multi-agent, multi-reasoning-path prompting for complex problem solving},
  author={Chen, Pei and Zhang, Shuai and Han, Boran},
  booktitle={Findings of the Association for Computational Linguistics: NAACL 2024},
  pages={1720--1738},
  year={2024}
}

@article{liu2024marscode,
  title={Marscode agent: Ai-native automated bug fixing},
  author={Liu, Yizhou and Gao, Pengfei and Wang, Xinchen and Liu, Jie and Shi, Yexuan and Zhang, Zhao and Peng, Chao},
  journal={arXiv preprint arXiv:2409.00899},
  year={2024}
}

@article{gao2025trae,
  title={Trae agent: An llm-based agent for software engineering with test-time scaling},
  author={Gao, Pengfei and Tian, Zhao and Meng, Xiangxin and Wang, Xinchen and Hu, Ruida and Xiao, Yuanan and Liu, Yizhou and Zhang, Zhao and Chen, Junjie and Gao, Cuiyun and others},
  journal={arXiv preprint arXiv:2507.23370},
  year={2025}
}

@article{yang2025lingxi,
  title={Lingxi: Repository-Level Issue Resolution Framework Enhanced by Procedural Knowledge Guided Scaling},
  author={Yang, Xu and Zhou, Jiayuan and Pacheco, Michael and Zhu, Wenhan and He, Pengfei and Wang, Shaowei and Liu, Kui and Pan, Ruiqi},
  journal={arXiv preprint arXiv:2510.11838},
  year={2025}
}

@article{xue2025pagent,
  title={PAGENT: Learning to Patch Software Engineering Agents},
  author={Xue, Haoran and Uddin, Gias and Wang, Song},
  journal={arXiv preprint arXiv:2506.17772},
  year={2025}
}

@inproceedings{yao2022react,
  title={React: Synergizing reasoning and acting in language models},
  author={Yao, Shunyu and Zhao, Jeffrey and Yu, Dian and Du, Nan and Shafran, Izhak and Narasimhan, Karthik R and Cao, Yuan},
  booktitle={The eleventh international conference on learning representations},
  year={2022}
}

@article{kirk2023understanding,
  title={Understanding the effects of rlhf on llm generalisation and diversity},
  author={Kirk, Robert and Mediratta, Ishita and Nalmpantis, Christoforos and Luketina, Jelena and Hambro, Eric and Grefenstette, Edward and Raileanu, Roberta},
  journal={arXiv preprint arXiv:2310.06452},
  year={2023}
}

@article{jin2025rl,
  title={Rl is neither a panacea nor a mirage: Understanding supervised vs. reinforcement learning fine-tuning for llms},
  author={Jin, Hangzhan and Lv, Sicheng and Wu, Sifan and Hamdaqa, Mohammad},
  journal={arXiv preprint arXiv:2508.16546},
  year={2025}
}

@misc{pyright,
  author = {{Microsoft Corporation}},
  title = {Pyright: Static Type Checker for Python},
  year = {2024},
  publisher = {GitHub},
  journal = {GitHub repository},
  howpublished = {\url{https://github.com/microsoft/pyright}}
}

@article{wang2025swe,
  title={Swe-mirror: Scaling issue-resolving datasets by mirroring issues across repositories},
  author={Wang, Junhao and Zan, Daoguang and Xin, Shulin and Liu, Siyao and Wu, Yurong and Shen, Kai},
  journal={arXiv preprint arXiv:2509.08724},
  year={2025}
}

@article{liu2025graphlocator,
  title={GraphLocator: Graph-guided Causal Reasoning for Issue Localization},
  author={Liu, Wei and Peng, Chao and Gao, Pengfei and Liu, Aofan and Zhang, Wei and Zhao, Haiyan and Jin, Zhi},
  journal={arXiv preprint arXiv:2512.22469},
  year={2025}
}

@article{chen2025beyond,
  title={Beyond two-stage training: Cooperative sft and rl for llm reasoning},
  author={Chen, Liang and Han, Xueting and Shen, Li and Bai, Jing and Wong, Kam-Fai},
  journal={arXiv preprint arXiv:2509.06948},
  year={2025}
}

@article{chen2025synergy,
  title={The synergy dilemma of long-cot sft and rl: Investigating post-training techniques for reasoning vlms},
  author={Chen, Jierun and Yu, Tiezheng and Bai, Haoli and Yao, Lewei and Wu, Jiannan and Li, Kaican and Mi, Fei and Tao, Chaofan and Zhu, Lei and Zhang, Manyi and others},
  journal={arXiv preprint arXiv:2507.07562},
  year={2025}
}
\bibliographystyle{icml2026}

%%%%%%%%%%%%%%%%%%%%%%%%%%%%%%%%%%%%%%%%%%%%%%%%%%%%%%%%%%%%%%%%%%%%%%%%%%%%%%%
%%%%%%%%%%%%%%%%%%%%%%%%%%%%%%%%%%%%%%%%%%%%%%%%%%%%%%%%%%%%%%%%%%%%%%%%%%%%%%%
% APPENDIX
%%%%%%%%%%%%%%%%%%%%%%%%%%%%%%%%%%%%%%%%%%%%%%%%%%%%%%%%%%%%%%%%%%%%%%%%%%%%%%%
%%%%%%%%%%%%%%%%%%%%%%%%%%%%%%%%%%%%%%%%%%%%%%%%%%%%%%%%%%%%%%%%%%%%%%%%%%%%%%%
\newpage

\appendix
\onecolumn

% \usepackage{multirow}
% \usepackage{graphicx}
\begin{table}[]

\centering
\caption{Illustration of the tool sets of agentic methods (Agentless \cite{xia2024agentless} is not included). Among all methods, only RepoNavigator takes the \texttt{jump} logic into consideration, and RepoNavigator has the smallest number of tools.}
\resizebox{\columnwidth}{!}{%
\begin{tabular}{cclll}
\toprule
\multicolumn{1}{l}{\textbf{Method}}          & \multicolumn{1}{l}{\textbf{Number}} & \textbf{Tools}                         & \textbf{Input}                                                                    & \textbf{Output}                                                 \\
\midrule
\multirow{3}{*}{CoSIL}        & \multirow{3}{*}{3}         & search\_class\_node           & file\_path, class\_name                                                  & The code snippet of a class node.                      \\
                              &                            & search\_class\_function\_node & file\_path, class\_name, function\_name                                  & Get the code snippet of a class member function node.  \\
                              &                            & search\_file\_function\_node  & file\_path, function\_name                                               & Get the code snippet of a static function node.        \\
\midrule
\multirow{3}{*}{LocAgent}     & \multirow{3}{*}{3}         & SearchEntity                  & Keywords                                                                 & Related entities with code snippets.                   \\
                              &                            & TraverseGraph                 & Start entity ids, direction, traverse hops, entity types, relation types & A traversed subgraph with entities and relations.      \\
                              &                            & RetrieveEntity                & Entity ids                                                               & The complete code of a specific entity.                \\
\midrule
\multirow{5}{*}{OrcaLoca}     & \multirow{5}{*}{5}         & search\_file\_contents        & file\_name, directory\_path                                              & The file contents or skeleton (if exceeds 200 lines).  \\
                              &                            & search\_class                 & class\_name, method\_name                                                & The class contents or skeleton (if exceeds 200 lines). \\
                              &                            & search\_method\_in\_class     & class\_name, method\_name, file\_path                                    & The method code snippet.                               \\
                              &                            & search\_callable              & query\_name, file\_path                                                  & The matched code snippet.                              \\
                              &                            & search\_source\_code          & file\_path, source\_code                                                 & The related function/class code snippet.               \\
\midrule
\multirow{5}{*}{RepoSearcher} & \multirow{5}{*}{5}         & GetRepoStructure              & None                                                                     & The repository file structure.                         \\
                              &                            & GetImportOfFile               & file, class                                                              & The imports of file.                                   \\
                              &                            & SearchClass                   & file, class                                                              & Code content of the searched class.                    \\
                              &                            & SearchFunction                & file, function                                                           & Code content of the searched function.                 \\
                              &                            & SearchClassMethod             & file, class, method                                                      & Code content of the searched method.                   \\
                              
\midrule
\rowcolor{blue!10}
Ours                          & 1                          & Jump                          & file\_path, symbol, index (optional)                                     & Definition code snippet of the symbol.       \\
\bottomrule
\end{tabular}%
}
\label{tab:toolset}
\end{table}

\section{Detailed Illustration of Baselines}
\label{sec:baselines}
The comprehensive toolsets for all aforementioned methods are detailed in Table~\ref{tab:toolset}. Furthermore, this section provides an in-depth discussion regarding the degree of autonomy exhibited by these approaches.
\paragraph{Agentless} 
Agentless \cite{xia2024agentless} is a workflow for issue localization. First, it identifies suspicious files in the repository. Second, relevant classes and functions are detected. Third, precise locations for editing are given by LLMs based on the classes and functions.
\paragraph{CoSIL}
CoSIL \cite{jiang2025cosil} is an agent that first conducts file-level localization and then conducts function-level localization. CoSIL dynamically constructs call graphs of modules (classes, functions) during the repo-level searching process, and applies context pruning to effectively reduce the searching scope.
\paragraph{LocAgent}
LocAgent \cite{chen-etal-2025-locagent} is almost a fully-automatic LLM agent, besides its planning prompt concatenated into the context at the beginning of the searching process. It builds the whole repository into a direct heterogeneous graph, whose nodes are files, classes, and functions. Additionally, edges are built by dependencies such as imports and invocations. Multiple graph-level searching tools are equipped to the LLM for multi-hop reasoning.
\paragraph{OrcaLoca}
OrcaLoca \cite{yu2025orcalocallmagentframework} is an LLM-based agent framework for software issue localization that performs repository-level exploration via priority-based action scheduling. It incrementally decomposes file- and class-level searches into fine-grained function-level actions with relevance scoring, while applying distance-aware context pruning on a code graph to efficiently narrow down potential bug locations. Because we failed to reproduce the code of OrcaLoca, we did not evaluate it on SWE-ben\_Pro, and we directly report the results (provided by \citet{jiang2025cosil}) on SWE-bench\_Verified.
\paragraph{RepoSearcher}
RepoSearcher \cite{ma2025toolintegratedreinforcementlearningrepo} is an agent that first conducts file-level localization and then function-level localization, which aligns with CoSIL. RepoSearcher introduced the first training framework \textit{ToolTrain} for localization agents, which is composed of distilling from a closed-source model (Claude3.7-Sonnet in Reposearcher) as warmup and reinforcement learning to further enhance the performance.
\paragraph{Ours}
Compared with all baselines, we are the first fully-automatic LLM agent, with no fixed workflow and no planning prompt, and we are the first method trained directly from pretrained open-source LLMs without a closed-source teacher model. Lastly, we only integrate a single yet powerful tool to the agent, which reduces compounding error and narrows the access scope of the agent.

\section{Experimental Details}
\label{sec:more_details}
\paragraph{Benchmarks}
The SWE-bench \cite{jimenez2023swe} dataset series is designed to evaluate the capability of LLM agents in resolving real-world GitHub issues. In the official evaluation protocol, models are provided solely with the problem statement and the historical repository state to generate a patch. However, in practical software engineering workflows, developers typically have access to at least one entry point for any given issue. To better simulate these real-world scenarios and enhance evaluation fidelity, we deviate from the official SWE-bench standard by providing an explicit entry point for each sample. Specifically, for both our proposed method and all baseline models, we supply the precise file path and the corresponding entry function.
\paragraph{Tools Description} Table~\ref{tab:toolset} illustrates the detailed input and output of RepoNavigator and baselines (except Agentless, which is workflow-based). CoSIL, OrcaLoca, and RepoSearcher apply simple retrieval tools that overlook the structural information of the repository. Among all methods, only RepoNavigator reduces the number of tools to \textbf{one}.
\paragraph{Hyperparameters for RL}
We set \texttt{clip\_ratio\_low} to 0.2, \texttt{clip\_ratio\_high} to 0.8, learning rate to $10^{-6}$, \texttt{training\_batch\_size} to 128, \texttt{temperature} to 1.0, \texttt{maximum\_toolcalling} to 12, and \texttt{max\_response\_length} to 10240. 
\paragraph{Hyperparameters for Baselines}
For baseline methods, we use vLLM \cite{kwon2023efficient} to deploy the corresponding model locally, and we set \texttt{max\_response\_length} to 32768, \texttt{enforce-eager} to True. We deploy all models on 8 NVIDIA Tesla-A100-80G GPUs. We follow the default settings of vLLM for other hyperparameters.
\paragraph{Metrics}
\label{sec:metrics}
Given the set of predicted locations (either file-level or function-level) \(\hat{Y}\), and the set of groundtruth locations \(Y^*\), the aforementioned metrics are calculated as:
\begin{equation}
    \text{Recall}=\frac{|\hat{Y}\cap Y^*|}{|Y^*|}
\end{equation}
\begin{equation}
    \text{Precision}=\frac{|\hat{Y}\cap Y^*|}{|\hat{Y}|}
\end{equation}
\begin{equation}
    \text{Sample-F1}=\frac{2\times |\hat{Y}\cap Y^*|}{|\hat{Y}|+ |Y^*|}
\end{equation}
\begin{equation}
    \text{IoU}=\frac{|\hat{Y}\cap Y^*|}{|\hat{Y}\cup Y^*|}
\end{equation}
In practice, when the prediction set \(\hat{Y}\) is empty (for instance, total failure), we set recall, precision, sample-F1, and IoU to zero. We use the function-level localization result of different methods and apply the patch generation backend in Agentless \cite{xia2024agentless} to generate patches. Resolved(\%) denotes the percentage of samples that pass all test units after applying the patch in Table~\ref{tab:resolve}.
\begin{table*}[t]
\centering
\caption{Comparison of tool configurations for RepoNavigator. \cmark \space denotes having the tool, and \xmark \space denotes not having the tool. When \texttt{Jump} is equipped, excessive tools do not improve the performance.}
\begin{tabular}{cccc|cccc}

\toprule
\multirow{2}{*}{\textbf{Jump}} & \multirow{2}{*}{\textbf{GetClass}} & \multirow{2}{*}{\textbf{GetFunc}} & \multirow{2}{*}{\textbf{GetStruc}} &
\multicolumn{4}{c}{\textbf{Function-level}} \\
\cmidrule(lr){5-8}
& & & & Recall & Precision & Sample-F1 & IoU \\
\midrule
\cmark & \cmark & \cmark & \cmark & 14.28 & 15.44 & 14.40 & 13.71 \\
\cmark & \cmark & \cmark & \xmark & 22.60 & 25.02 & 22.80 & 21.44 \\
\cmark & \xmark & \xmark & \cmark & 24.64 & 27.48 & 25.05 & 24.00 \\
\rowcolor{blue!10}
\cmark & \xmark & \xmark & \xmark & \textbf{25.11} & \textbf{29.16} & \textbf{25.75} & \textbf{24.28} \\
\bottomrule
\end{tabular}

\label{tab:m_vs_s_full}
\end{table*}

\paragraph{Implementation}
When the response exceeds the maximum length, we clip and force the agent to stop, and we give zero as its score. When the agent exceeds the maximum tool-calling times (which is 12), we add \textbf{``You must not call tools anymore, and you must give the final answer"} to the tool's response. Most of the time, the agent will stop calling tools and generate the final response. If not, we force it to stop and give zero as its score. Note that when the maximum tool-calling times are not achieved, and the final answer is generated, the agent loop will stop automatically. The aforementioned process is an automatic agentic framework, which allows the agent to explore environments with few constraints. 

\paragraph{Language Server}
In practice, we apply a Python language server, Pyright \cite{pyright}, to extract the definition code corresponding to an invoked symbol within a repository. However, the presence of monkey patches—runtime modifications to the repository—and dynamic imports can degrade the performance of the language server, as its functionality relies on static analysis techniques such as abstract syntax trees and symbol tables. When such circumstances occur, the tool returns an error message indicating that the definition of the current symbol cannot be located due to unknown reasons. Nevertheless, in our empirical evaluation, we did not observe any instances of monkey patching or dynamic imports within the analyzed datasets.
\section{More Results}
\subsection{Ablation Study}
\label{sec:ablation_study}
We investigate the impact of the hybrid reward mechanism in Fig. \ref{fig:ablation_study} and explore the influence of SFT as a ``cold start" phase for the Reinforcement Learning (RL) process. As illustrated in Table \ref{tab:ablation_study}, our findings suggest that while SFT enhances initial model performance, it ultimately constrains the long-term potential of RL. By forcing the model to adhere to pre-generated trajectories, SFT improves basic task proficiency at the expense of autonomous exploration. This finding aligns with previous research on this topic \cite{chen2025beyond,jin2025rl,chen2025synergy}. Consequently, the results in Table \ref{tab:ablation_study} demonstrate that training the model directly with reinforcement learning yields the most superior performance.
\subsection{Cost Analysis}
We conduct a comparative analysis of token usage across multiple methodologies, utilizing Qwen2.5-14B-Instruct as the base model on SWE-bench\_Pro. Empirical results (see Table~\ref{tab:cost_analysis}) indicate that RepoNavigator represents a superior agentic design in terms of cost-efficiency. While baseline approaches such as CoSIL and RepoSearcher incur substantial overhead by indexing all repository files—leading to significantly higher average input tokens—our proposed method optimizes token consumption by avoiding such exhaustive operations.
\begin{table*}[!t]
\centering
\caption{The model achieves competitive and steady scores as \(\lambda\) ranges from 0.4 to 0.6. This insensitivity suggests that the heuristic setting of \(\lambda=0.5\) is near-optimal and the synergy between outcome and tool rewards is inherently stable, rather than being sensitive to specific weighting.}
\label{tab:hyper}
\small
\setlength{\tabcolsep}{5pt}
\renewcommand{\arraystretch}{1.2}
\begin{tabular}{c|cccc|cccc}

\toprule
% \multirow{2}{*}{\textbf{Model}} & 
\multirow{2}{*}{\textbf{\(\lambda\)}}  & 
\multicolumn{4}{c|}{\textbf{Function-level}} &
\multicolumn{4}{c}{\textbf{File-level}} \\
\cmidrule(lr){2-5} \cmidrule(lr){6-9}
  & {Recall} & {Precision} & {Sample-F1} & {IoU} 
& {Recall} & {Precision} & {Sample-F1} & {IoU} \\

\midrule
\midrule
 0.6 & 26.32 & 28.86 & 26.85 & 25.88 & 50.93 & 52.66 & 51.18 & 50.23\\
 0.5 & 26.69 & 30.34 & 27.49 & 26.43 & 50.62 & 53.83 & 51.63 & 50.62\\
 0.4 & 27.18 & 30.70 & 28.08 & 27.18 & 50.67 & 52.56 & 51.49 & 50.67\\
 \bottomrule
\end{tabular}
\end{table*}
\begin{table}[!t]
\centering

\label{tab:ablation_side_by_side}

% 左侧表格

\begin{minipage}{0.48\textwidth}
\centering
% \small % 略微缩小字体以防溢出
\caption{Ablation study on SFT and RL.}
\begin{tabular}{cc|cccc}
\toprule
\multirow{2}{*}{\textbf{SFT}} & \multirow{2}{*}{\textbf{RL}} & \multicolumn{4}{c}{\textbf{Function-level}} \\
\cmidrule(lr){3-6}
& & Recall & Precision & Sample-F1 & IoU \\
\midrule
\xmark & \xmark & 15.89 & 17.46 & 16.19 & 15.46 \\
\cmark & \xmark & 24.92 & 27.44 & 25.38 & 24.01 \\
\cmark & \cmark & 25.93 & 30.08 & 26.97 & 25.75 \\
\rowcolor{blue!10}
\xmark & \cmark & \textbf{26.69} & \textbf{30.34} & \textbf{27.49} & \textbf{26.43} \\
\bottomrule
\end{tabular}
\label{tab:ablation_study}
\end{minipage}
\hfill % 填充中间空白
% 右侧表格
\begin{minipage}{0.48\textwidth}
\centering

\caption{Cost Analysis.}
\begin{tabular}{c|cccc}
\toprule
\multirow{2}{*}{\textbf{Method}} & \multicolumn{3}{c}{\textbf{Avg. Tokens (\(\times10^3\))}} \\
\cmidrule(lr){2-4}
&  Input & Output & Total  \\
\midrule
 
 Agentless & 19.84 & 2.36 & 22.20 \\
 CoSIL & 7.93 & 0.50 & 8.44 \\
 LocAgent & 45.88 & 2.98 & 48.86 \\
 RepoSearcher & 23.85 & 2.81 & 26.66\\
 % RepoNavigator & 25.93 & 30.08 & 26.97\\
\rowcolor{blue!10}
 RepoNavigator & 1.73 & 0.95 & 2.68  \\
\bottomrule
\end{tabular}
\label{tab:cost_analysis}
\end{minipage}
\end{table}
\subsection{Rationality of the Joint Reward Mechanism.}
\begin{figure}
    \centering
    \includegraphics[width=0.5\linewidth]{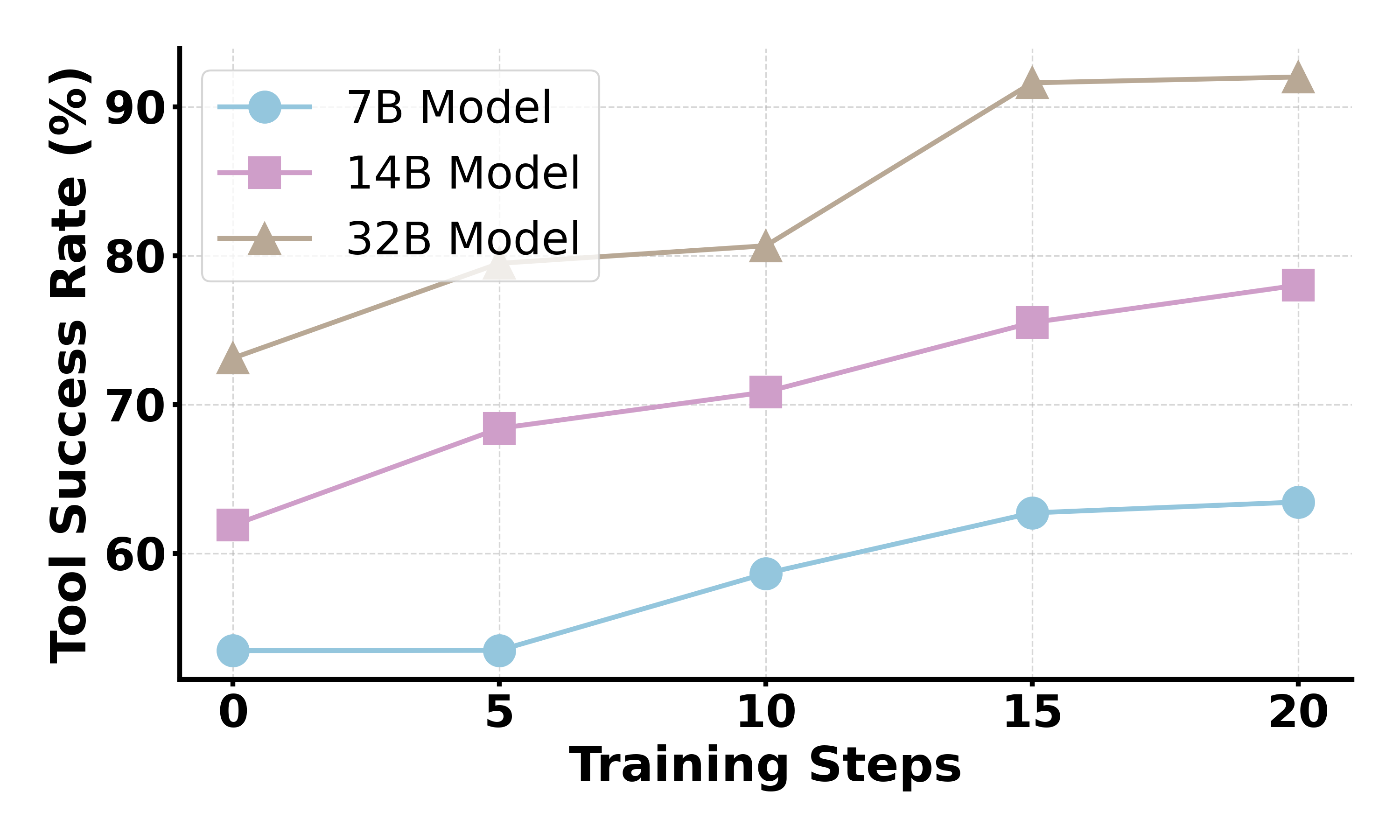}
    \caption{Tool success rate of the 7B, 14B, and 32B model.}
    \label{fig:tool_reward}
\end{figure}

\begin{figure}[!t]
    \centering
    \includegraphics[width=0.5\linewidth]{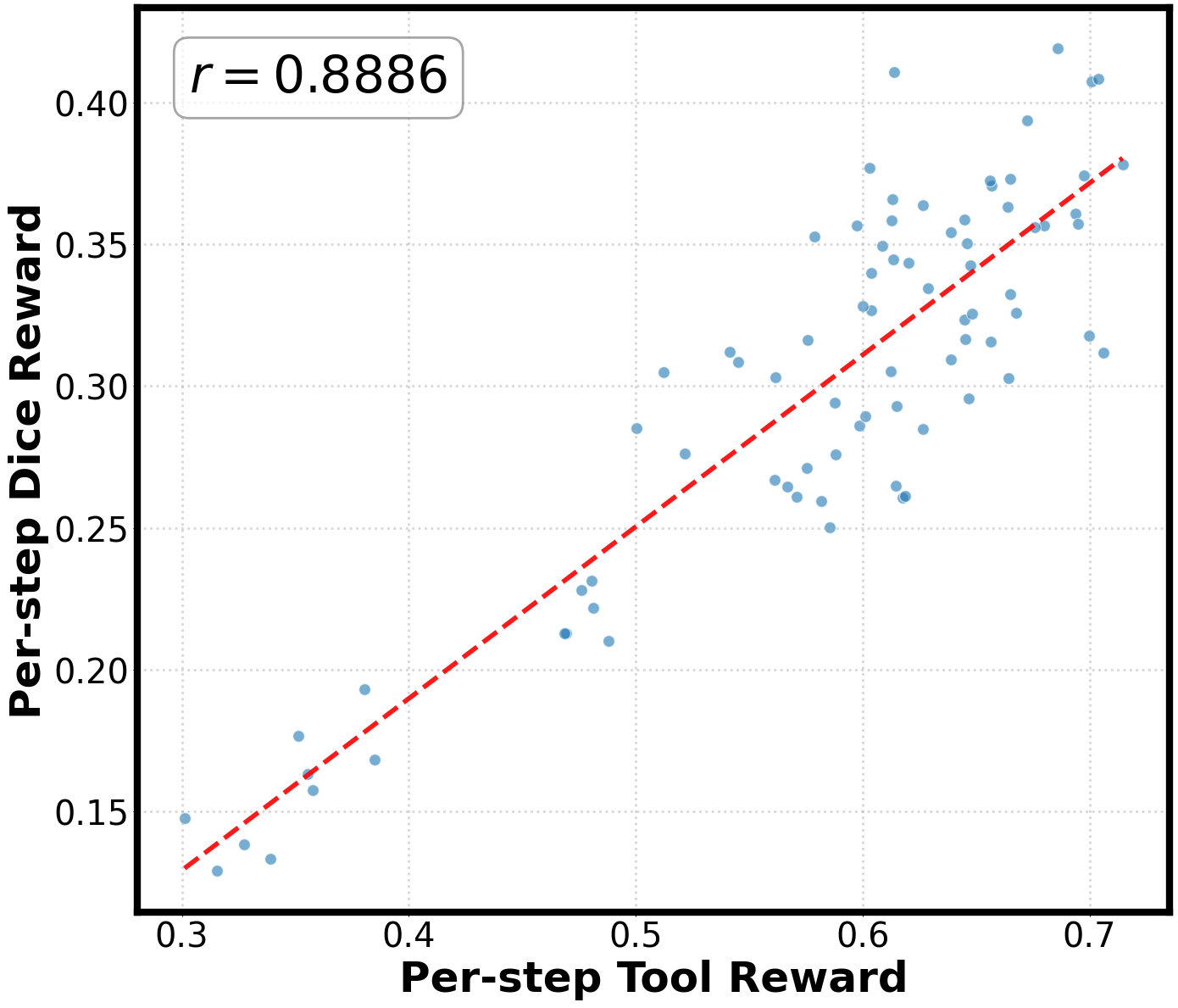}
    \caption{Evolution of the average tool reward during the training process. Higher rewards indicate improved efficiency in selecting and executing external tools.}
    \label{fig:correlation}
\end{figure}

 The integration of tool reward and dice reward as a hybrid reinforcement learning objective is grounded in two key empirical observations. First, as illustrated in Fig.~\ref{fig:tool_reward}, the tool reward exhibits a consistent and stable upward trajectory during the training phase, demonstrating that the agent is effectively learning to master tool-use patterns and that this reward signal provides a dense, optimizable gradient. Second, Fig.~\ref{fig:correlation} reveals a high Pearson correlation (r=0.8886) between tool reward and dice reward. This strong correlation indicates that the intermediate tool-use proficiency is a reliable proxy for the final task success (as measured by the Dice score). By combining these two signals, the reward function not only captures the ultimate goal but also provides informative "stepping stones" during training, effectively mitigating the sparse reward problem while ensuring that the optimization of tool interaction aligns to improve overall performance.
 \subsection{Sensitivity Analysis}
  \label{sec:hyperparameter}
In the design of our agent, we adopt a balanced heuristic by assigning equal weight to both outcome and tool rewards (i.e., a 1:1 ratio). This choice represents a non-informative prior, assuming that the correctness of the final result and the adherence to tool protocols are of equal importance. To evaluate the robustness of our method, we conduct a sensitivity study on the reward weighting factor \(\lambda\). As illustrated in Table~\ref{tab:hyper}, the performance on the test set remains remarkably stable as \(\lambda\) varies.

% \paragraph{Evolution of tool-calling proficiency}
% We investigate the impact of the hybrid reward mechanism in Fig. \ref{fig:ablation_study} and explore the influence of SFT as a ``cold start" phase for the Reinforcement Learning (RL) process. As illustrated in Table \ref{tab:ablation_study}, our findings suggest that while SFT enhances initial model performance, it ultimately constrains the long-term potential of RL. By forcing the model to adhere to pre-generated trajectories, SFT improves basic task proficiency at the expense of autonomous exploration. Consequently, the results in Table \ref{tab:ablation_study} demonstrate that training the model directly with reinforcement learning yields the most superior performance.
% We evaluate the tool-calling performance during training in Fig.~\ref{fig:tool}, where tool rewards consistently improve. A potential side effect of integrating success rates into the GRPO reward function is the emergence of conservative" agents that inflate success metrics by suppressing tool usage. However, the results in Fig.~\ref{fig:turns} and Fig.~\ref{fig:succ_turns} mitigate this concern, showing that successful tool-calling turns increase while overall turns remain stable or rise. We observe a particularly pronounced increase in total turns for the 7B model (Fig.~\ref{fig:turns}). Unlike the 14B and 32B models, which leverage their superior reasoning to reach targets with fewer, more precise invocations, the 7B model undergoes a more intensive exploration phase due to its initially lower tool-mastery (Fig.~\ref{fig:tool_rewards}).
\section{Detailed Analysis}
\subsection{\texttt{Jump} as a Learnable Action Abstraction}
\label{sec:theory_jump}

We provide an analysis to explain why the proposed \texttt{Jump} tool constitutes a learnable and effective action abstraction for repository-level code navigation. We use Python as an example; other languages that are similar to Python (C++, Java...) are also suitable.

\paragraph{Problem formulation.}
We model repository navigation as a finite-horizon Markov Decision Process (MDP)
$\mathcal{M} = (\mathcal{S}, \mathcal{A}, P, R)$,
where a state $s \in \mathcal{S}$ represents the agent’s current context (including all historical conversations),
and an action $a \in \mathcal{A}$ corresponds to invoking a navigation tool.
In prior approaches, the action space $\mathcal{A}$ is composed of multiple heterogeneous tools (e.g., SearchClass, SearchFunction, GetImports...), each inducing distinct and noisy transitions.
This results in a large effective branching factor and long-horizon exploration difficulty.

\paragraph{Action abstraction via \texttt{Jump}.}
Our method replaces the heterogeneous action space $\mathcal{A}$ with a single abstract action space
$\mathcal{A}' = \{\texttt{Jump}(s)\}$,
where each action deterministically resolves a symbol to its definition using static analysis.
Formally, this defines an abstraction mapping:
\[
\phi: \mathcal{A} \rightarrow \mathcal{A}',
\]
which collapses multiple low-level navigation actions into a single execution-aware transition.
Importantly, this abstraction is \emph{many-to-one} and preserves execution-relevant semantics, as symbol resolution directly reflects Python’s name-binding rules.

\paragraph{Preservation of Reachability.}
Although \texttt{Jump} does not have access to oracle ground-truth locations, it preserves the reachability of task-relevant program elements under realistic assumptions.
Specifically, if a target symbol is statically resolvable within the repository, then there exists a sequence of \texttt{Jump} actions that reaches its definition.
Consequently, the abstraction does not remove optimal solutions, but instead restricts exploration to semantically valid transitions.

\paragraph{Reduction of Exploration Complexity.}
The primary benefit of the abstraction lies in its effect on exploration.
Let $B$ denote the average branching factor induced by the original action space $\mathcal{A}$, and $B'$ that induced by $\mathcal{A}'$.
Because \texttt{Jump} deterministically resolves symbols and filters out irrelevant candidates, we empirically observe that:
\[
\mathbb{E}[B'] \ll \mathbb{E}[B].
\]
In long-horizon tasks, the probability of entering irrecoverable exploration states grows exponentially with the branching factor.
Thus, reducing the branching factor substantially improves the tractability of reinforcement learning under finite rollouts.

\paragraph{Implications for Policy Learning.}
The abstraction reshapes the learning problem by aligning actions with the underlying program structure.
As a result, the induced MDP $\mathcal{M}' = (\mathcal{S}, \mathcal{A}', P', R)$ admits policies that are easier to optimize, even with standard policy-gradient methods.
This explains why reinforcement learning from a pretrained language model succeeds in our setting without reliance on distillation from closed-source teacher LLMs.

% \paragraph{Empirical alignment.}
% Our analysis is supported by empirical results in Fig.~\ref{fig:tas}, where \texttt{Jump} achieves significantly higher overlap between accessed scopes and ground-truth solution regions.
% This indicates that the abstraction not only reduces the search space, but also concentrates exploration on semantically relevant program locations.

In summary, \texttt{Jump} is not merely a tool simplification, but a principled action abstraction that preserves reachability while dramatically reducing exploration complexity, thereby enabling effective reinforcement learning for repository-level software engineering tasks.

\subsection{Necessity of Reinforcement Learning for Long-Horizon Repository Navigation}
The objective of the agent is not merely to invoke the jump operation correctly, but to learn a policy for sequencing execution-level navigation steps under sparse terminal rewards. Intermediate states do not admit dense supervision, and optimal behavior depends on long-horizon credit assignment.
Reinforcement learning provides a principled framework for optimizing such policies, enabling the agent to balance exploration depth, termination decisions, and recovery from incorrect jumps. In contrast, supervised fine-tuning would require trajectory-level annotations or strong heuristics to approximate long-term returns, which are unavailable in this setting.
\subsection{Language-agnostic Implementation}
RepoNavigator is designed to be language-agnostic, relying only on the availability of execution-level navigation primitives provided by a language server, such as symbol resolution or call-site traversal. These primitives are supported across a wide range of programming languages through standard language server protocols.
In our experiments, we focus on Python repositories to simplify implementation and evaluation. This choice is made for experimental convenience rather than methodological necessity. The underlying formulation and learning framework do not depend on Python-specific semantics, and can be directly instantiated for other languages given corresponding language server support.
\subsection{Action Space Reduction via a Single Tool}
Restricting the agent to a single execution-aware navigation primitive substantially reduces the branching factor of the decision process. For long-horizon tasks with sparse terminal rewards, the probability of success of a trajectory scales multiplicatively with per-step correctness. Expanding the action space with multiple tools increases action entropy and degrades this multiplicative success rate unless strong priors are available. Our empirical results suggest that, in repository-level reasoning, depth of execution-aware exploration is more critical than breadth of tool diversity.
\subsection{Preventing Data Leakage}
A common concern is whether pre-training data leakage could undermine the validity of post-training methods. We mitigate this concern through both dataset curation and temporal separation. Specifically, the SWE-bench\_Pro dataset was released in 2025, whereas the Qwen2.5 model family was published in 2024, making direct pre-training exposure unlikely. In addition, we explicitly remove any training samples whose repositories overlap with those in SWE-bench\_Verified or SWE-bench\_Pro.
Together, these measures substantially reduce the risk that the observed performance gains arise from data leakage rather than the proposed training procedure.

\section{More Related Works}
To enable efficient training, some methods have proposed a simulated environment for agents to interact with. SWE-Gym \cite{pan2024training} proposed the first Gym environment to train SWE agents. SWE-mirror \cite{wang2025swe} mirrors issues across datasets to scale up. To stablize the performance, we use the large-scale training datasets of SWE-smith, which is conservative but has stable performance.

Aside from baseline methods we reproduced in our research, there are also agentic designs for repo-level tasks. PAGENT \cite{xue2025pagent} analyzes the failure reasons of existing agents and applies CFG creation and exploration to infer the type of a patch. MarsCode \cite{liu2024marscode} is a training-free multi-agent framework including a manager, a searcher, a tester, a reproducer, and editors.  Trae \cite{gao2025trae} is a training-free workflow composed of patch generation, regression test, patch pruning, and selection. Among all agentic frameworks, Trae has the highest rank on the SWE-bench leaderboard. Linxi \cite{yang2025lingxi} is a training-free multi-agent framework, while adding the repository as a knowledge database, and retrieval-based methods are applied to extract knowledge from this database. PatchPilot \cite{li2025patchpilot} adds a reproduction and refinement process in the workflow. CORTEXA \cite{sohrabizadehnemotron} reflects the repository into an embedding space to enable efficient retrieval. \citet{chen-etal-2025-locagent} builds the repository as a graph and applies graph-level searching tools for localization, and \citet{wang2025openhands} furthermore integrates commit history as agent memory. RepoLens \cite{wang2025extracting} equips conceptual information of the repository to enable repo-level understanding. These pipelines are training-free and yield competitive results. These workflows are sophisticated, and they successfully integrate repo-level tools, including a language server, terminal, and search engine. However, their complexity hinders their potential for training. On the other hand, we remove unnecessary tools and workflows, preserving only the \texttt{jump} tool (built upon the language server) to build a fully-automatic agent. Our design enables efficient training with reinforcement learning.
\section{Illustration of Training}
\label{sec:algorithm}
We present the detailed pseudo code of the RL training process of RepoNavigator in Algorithm~\ref {alg:reponavigator}. For simplicity, we include the entry point (normally a testing function) in issue \(q\).
\begin{algorithm}[!t]
\caption{Reinforcement learning process of RepoNavigator using GRPO}
\label{alg:reponavigator}
\begin{algorithmic}[1]
\REQUIRE Repository $\mathcal{R}$, issue $q$, initial policy $\pi_\theta$, group size $G$
\ENSURE Output $\hat{Y}$ maximizing $\mathbb{E}_{\tau \sim \pi_\theta}[R(\tau)]$

\STATE $o_0 \gets q$, $\tau \gets \emptyset$

\WHILE{not terminated}
    \STATE $h_t \gets (q, o_{1:t-1}, a_{1:t-1})$
    \STATE \textit{Optionally generate intermediate reasoning}
    \STATE $x_t \sim \pi_\theta(\cdot \mid h_t)$
    
    \IF{$x_t$ is a \textbf{tool-call}}
        \STATE $a_t \gets x_t$
        \STATE $o_t \gets T(a_t, \mathcal{R})$
        \STATE $\tau \gets \tau \cup \{(h_t, a_t)\}$
    \ELSE
        \STATE $\hat{Y} \gets x_t$
        \STATE \textbf{break}
    \ENDIF
\ENDWHILE

\STATE Sample a group of trajectories $\{\tau^{(g)}, \hat{Y}^{(g)}\}_{g=1}^{G}$ using $\pi_\theta$
\STATE Compute rewards $\{R^{(g)} = R(\hat{Y}^{(g)}, Y^*, \tau^{(g)})\}_{g=1}^{G}$

\STATE Compute group baseline $\bar{R} \gets \frac{1}{G} \sum_{g=1}^{G} R^{(g)}$
\FOR{each trajectory $\tau^{(g)}$}
    \STATE $\hat{A}^{(g)} \gets R^{(g)} - \bar{R}$
    \FOR{each $(h_t, a_t) \in \tau^{(g)}$}
        \STATE Assign advantage $\hat{A}_t \gets \hat{A}^{(g)}$
    \ENDFOR
\ENDFOR

\STATE Estimate GRPO loss $\mathcal{L}^{\text{GRPO}}(\theta)$ using Eq.~\ref{eq:grpo}

\end{algorithmic}
\end{algorithm}

\newpage
\section{Threats to Validity}
\paragraph{Groundtruth Retrieval}
A limitation of our work lies in the extraction of groundtruth locations. We extract modified locations directly from the \texttt{gold\_patch} in the datasets, which may ignore other patches that also resolve the issue. Our evaluation metrics do not consider these correct alternatives. However, using golden patches is acceptable when comparing multiple methods. If a method reveals golden locations (locations in golden patches), it undoubtedly contributes to the resolution of the issue, and the result in Table~\ref{tab:resolve} demonstrates this claim.

\paragraph{Language Limit}
Another limitation is that we only evaluate Python repositories in our experiments. This is because each language (C/C++, Java, etc.) has its unique language server, and we only succeeded in implementing the language server of Python. We will implement more language servers and validate our approach on more programming languages in the future.
 
\section{Case Study}
In this section, we present the full trajectory of RepoNavigator on \textit{astropy\_astropy-12907} from SWE-bench\_Verified in Fig. 7. We apply the default tool-calling prompt template of verl \cite{shen2024llm} and present an example. The system prompt and user prompt are presented in Fig. 7. Note that we do not present any process restrictions in our prompt, encouraging RepoNavigator to plan, call tools, and make decisions fully automatically. This is distinct from Agentless (which has a fixed workflow), LocAgent (which predefines a specific step-by-step workflow in its system prompt), CoSIL, and RepoSearcher (which are half-automatic because some forced steps are added to the workflow besides the automatic multi-turn tool-calling conversations).\\

Figure 7: The following pages are the prompt and a case study on \textit{astropy\_astropy\_12907} of SWE-bench\_Verified.
% \begin{figure*}[hbtp]
% \centering
% \begin{userbox}[User-turn-0]

% \begin{tcblisting}{
%   listing options={
%     basicstyle=\ttfamily\small,
%     columns=fullflexible,
%     keepspaces=true
%   },
%   breakable
% }
% ...(system prompt and problem statement are omitted)...

% The entry file of the code base is:

% astropy/modeling/tests/test_separable.py:

% def test_separable(compound_model, result):
%     assert_allclose(is_separable(compound_model), result[0])
%     assert_allclose(separability_matrix(compound_model), result[1])

% Your final answer should be all functions that should be modified, with the format of:

% relevant/path/to/file1.py::func_name1,relevant/path/to/file2.py::func_name2,...

% (a series of file::function separated by comma)

% Please put your final answer inside boxed{} only in the last turn.
% \end{tcblisting}

% \end{userbox}
% \end{figure*}

\begin{figure*}[hbtp]

    \centering
\begin{tcolorbox}[colback=red!5!white, colframe=red!75!black, title=Prompt]
\begin{verbatim}

[system]
You are Qwen, created by Alibaba Cloud. You are a helpful assistant.

# Tools

You may call one or more functions to assist with the user query.

You are provided with function signatures within <tools></tools> XML tags:
<tools>
{"type": "function", "function": {"name": "jump", "description": "In the 
specific file path, a symbol is referred and this tool can find 
where the tool is defined.
For instance, in the first turn, file_path is the entry point of.", 
"parameters": {"properties": {"symbol": {"description": "The symbol whose 
definition code will be given to the agent.", "type": "string"}, "file_path": 
{"description": "The relevant path to the file where the symbol is referred.", 
"type": "string"}}, "required": ["symbol", "file_path"], "type": "object"}}}
</tools>

For each function call, return a json object with function name and arguments 
within <tool_call></tool_call> XML tags:
<tool_call>
{"name": <function-name>, "arguments": <args-json-object>}
</tool_call>
[user]
You are given a codebase and an issue, you need to locate the files and 
functions causing this issue.
You can call the tool to check the definition code of a symbol. You can only 
check the symbol once for each turn.
The "file_path" is the relevant path of where the symbol is called, 
NOT where it is defined!
For instance, if "classA.functionB" is what you want to check (which is called 
in fileA.py), you should directly check "functionB" in "fileA.py".

This is the issue:
[Problem Statement]


The entry file of the code base is:
[Relevant Path To Entry Point]
[Entry Point]


Your final answer should be all functions that should be modified, such as: 
relevant/path/to/file1.py::func_name1,relevant/path/to/file2.py::func_name2,
...(a series of file::function pairs separated by comma)
Please put your final answer inside \boxed{} only in the last turn.
You can only call the tool once each turn.
For instance:
{"name": "jump", "arguments": {"symbol": "symbol_to_be_checked", "file_path": 
"file_where_the_symbol_is_used"}}
\end{verbatim}
\end{tcolorbox}
\label{fig:prompt}
\end{figure*}
% \begin{autowrapbox}
% def example\_function():

% for i in range(10):
%         print("This is a very very very very very very very very long line")
% \end{autowrapbox}

\begin{figure*}[hbtp]

    \centering
\begin{userbox}[User-turn-0]
\begin{verbatim}
    
...(system prompt and problem statement are omitted)...

The entry file of the code base is:
astropy/modeling/tests/test_separable.py:

def test_separable(compound_model, result):
    assert_allclose(is_separable(compound_model), result[0])
    assert_allclose(separability_matrix(compound_model), result[1])

Your final answer should be all functions that should be modified, 
with the format of: 

relevant/path/to/file1.py::func_name1,relevant/path/to/file2.py::func_name2,...
(a series of file::function separated by comma)

Please put your final answer inside \boxed{} only in the last turn.
\end{verbatim}

\end{userbox}
\label{fig:case_study}
\end{figure*}
\begin{figure*}[hbtp]
    \centering
\begin{llmbox}[LLM-turn-1]
\begin{verbatim}
<tool_call>
{"name": "jump", "arguments": {"symbol": "separability_matrix", "file_path": 
"astropy/modeling/tests/test_separable.py"}}
</tool_call>

\end{verbatim}
\end{llmbox}
\end{figure*}
\begin{figure*}[hbtp]
    \centering
\begin{userbox}[User-turn-2]
\begin{verbatim}
<tool_response>
The definition of symbol "separability_matrix" is:
def separability_matrix(transform):
    if transform.n_inputs == 1 and transform.n_outputs > 1:
        return np.ones((transform.n_outputs, transform.n_inputs), dtype=np.bool)
    separable_matrix = _separable(transform)
    separable_matrix = np.where(separable_matrix != 0, True, False)
    return separable_matrix


It is defined in: astropy/modeling/separable.py

You can check other symbols in previous files, or you can give the final answer.
</tool_response>
\end{verbatim}
\end{userbox}
\end{figure*}

\begin{figure*}[hbtp]
    \centering
\begin{llmbox}[LLM-turn-3]
\begin{verbatim}
The "separability_matrix" function is defined in "astropy/modeling/separable.py"
Given the issue with nested CompoundModels, it seems the problem might lie in 
how the function handles nested models. Specifically, the function might not 
correctly propagate the separability information through nested models.

To address this, we need to check the "_separable" function, which is likely 
where the separability logic is implemented. Let's check the "_separable" 
function.

<tool_call>
{"name": "jump", "arguments": {"symbol": "_separable", "file_path":
"astropy/modeling/separable.py"}}
</tool_call>

\end{verbatim}
\end{llmbox}
\end{figure*}
\begin{figure*}[hbtp]
    \centering
\begin{userbox}[User-turn-4]
\begin{verbatim}
<tool_response>
The definition of symbol "_separable" is:
def _separable(transform):
    """
        Calculate the separability of outputs.
        Parameters
        ----------
        transform : "astropy.modeling.Model"
        A transform (usually a compound model).
        Returns :
        is_separable : ndarray of dtype np.bool
        An array of shape (transform.n_outputs,) of boolean type
        Each element represents the separablity of the corresponding output.
    """
    if (transform_matrix := transform._calculate_separability_matrix()) 
        is not NotImplemented:
        return transform_matrix
    elif isinstance(transform, CompoundModel):
        sepleft = _separable(transform.left)
        sepright = _separable(transform.right)
        return _operators[transform.op](sepleft, sepright)
    elif isinstance(transform, Model):
        return _coord_matrix(transform, "left", transform.n_outputs)

It is defined in: astropy/modeling/separable.py

You can check other symbols in previous files, or you can give the final answer.
</tool_response>
\end{verbatim}
\end{userbox}
\end{figure*}
\begin{figure*}[hbtp]
    \centering
\begin{llmbox}[LLM-turn-5]
\begin{verbatim}
The "_separable" function is defined in "astropy/modeling/separable.py". This 
function is responsible for calculating the separability of outputs for a 
given model. Given the issue with nested CompoundModels, it seems the problem 
might lie in how the function handles the combination of separability matrices 
for nested models.

To address this, we need to ensure that the separability information is 
correctly propagated through nested models. Specifically, we need to check 
how "_operators" dictionary handles the combination of separability matrices.

Let's check the "_operators" dictionary.

<tool_call>
{"name": "jump", "arguments": {"symbol": "_operators", "file_path": 
"astropy/modeling/separable.py"}}
</tool_call>

\end{verbatim}
\end{llmbox}
\end{figure*}
\begin{figure*}[hbtp]
    \centering
\begin{userbox}[User-turn-6]
\begin{verbatim}
<tool_response>
The definition of symbol "_operators" is:
_operators = {"&": _cstack, "|": _cdot, "+": _arith_oper, "-": _arith_oper,
"*": _arith_oper, "/": _arith_oper, "**": _arith_oper}


It is defined in: astropy/modeling/separable.py

You can check other symbols in previous files, or you can give the final answer.
</tool_response>
\end{verbatim}
\end{userbox}
\end{figure*}
\begin{figure*}[hbtp]
    \centering
\begin{llmbox}[LLM-turn-7]
\begin{verbatim}
The "_operators" dictionary is defined in "astropy/modeling/separable.py". 
This dictionary maps operators to functions that handle the combination of 
separability matrices. Given the issue with nested CompoundModels, it seems 
the problem might lie in how the "_cstack" function handles the combination 
of separability matrices.

To address this, we need to check the "_cstack" function.

Let's check the "_cstack" function.

<tool_call>
{"name": "jump", "arguments": {"symbol": "_cstack", "file_path": 
"astropy/modeling/separable.py"}}
</tool_call>

\end{verbatim}
\end{llmbox}
\end{figure*}
\begin{figure*}[hbtp]
    \centering
\begin{userbox}[User-turn-8]
\begin{verbatim}
<tool_response>
The definition of symbol "_cstack" is:
def _cstack(left, right):
    noutp = _compute_n_outputs(left, right)
    
    if isinstance(left, Model):
    cleft = _coord_matrix(left, "left", noutp)
    else:
    cleft = np.zeros((noutp, left.shape[1]))
    cleft[: left.shape[0], : left.shape[1]] = left
    if isinstance(right, Model):
    cright = _coord_matrix(right, "right", noutp)
    else:
    cright = np.zeros((noutp, right.shape[1]))
    cright[-right.shape[0]:, -right.shape[1]:] = 1
    
    return np.hstack([cleft, cright])


It is defined in: astropy/modeling/separable.py

You can check other symbols in previous files, or you can give the final answer.
</tool_response>
\end{verbatim}
\end{userbox}
\end{figure*}
\begin{figure*}[hbtp]
    \centering
\begin{llmbox}[LLM-turn-9]
\begin{verbatim}
The "_cstack" function is defined in "astropy/modeling/separable.py". This 
function is responsible for handling the combination of separability matrices 
for the "&" operator. Given the issue with nested CompoundModels, it seems 
the problem might lie in how it handles the combination of separability
matrices for nested models.\n\nTo address this, we need to ensure that the 
separability information is correctly propagated through nested models. 
Specifically, we need to check how the "_coord_matrix" function handles 
the combination of separability matrices.

Let's check the "_coord_matrix" function.

<tool_call>
{"name": "jump", "arguments": {"symbol": "_coord_matrix", "file_path": 
"astropy/modeling/separable.py"}}
</tool_call>

\end{verbatim}
\end{llmbox}
\end{figure*}
\begin{figure*}[hbtp]
    \centering
\begin{userbox}[User-turn-10]
\begin{verbatim}
<tool_response>
The definition of symbol "_coord_matrix" is:
def _coord_matrix(model, pos, noutp):
    if isinstance(model, Mapping):
        axes = []
    for i in model.mapping:
        axis = np.zeros((model.n_inputs,))
        axis[i] = 1
        axes.append(axis)
        m = np.vstack(axes)
        mat = np.zeros((noutp, model.n_inputs))
    if pos == "left":
        mat[: model.n_outputs, :model.n_inputs] = m
    else:
        mat[-model.n_outputs:, -model.n_inputs:] = m
        return mat
    if not model.separable:
        # this does not work for more than 2 coordinates
        mat = np.zeros((noutp, model.n_inputs))
    if pos == "left":
        mat[:model.n_outputs, : model.n_inputs] = 1
    else:
        mat[-model.n_outputs:, -model.n_inputs:] = 1
    else:
        mat = np.zeros((noutp, model.n_inputs))
    for i in range(model.n_inputs):
        mat[i, i] = 1
        if pos == "right":
            mat = np.roll(mat, (noutp - model.n_outputs))
    return mat
    
    
It is defined in: astropy/modeling/separable.py

You can check other symbols in previous files, or you can give the final answer.
</tool_response>
\end{verbatim}
\end{userbox}
\end{figure*}
\begin{figure*}[hbtp]
    \centering
\begin{llmbox}[LLM-turn-11]
\begin{verbatim}
The "_coord_matrix" function is defined in "astropy/modeling/separable.py". This 
function is responsible for creating an array representing the inputs and 
outputs of a simple model. Given the issue with nested CompoundModels, 
it seems the problem might lie in how the function handles the combination of 
separability matrices for nested models.

To address this, we need to ensure that the separability information is 
correctly propagated through nested models. Specifically, we need to 
check how the function handles the combination of separability matrices 
for nested models. Based on the definitions provided, the issue seems to be 
with the "_cstack" function, which is responsible for handling the combination
of separability matrices for the "&" operator. The function does not correctly
handle the combination of separability matrices for nested models. 
Therefore, the functions that should be modified are:

\\boxed{astropy/modeling/separable.py::_cstack}

\end{verbatim}
\end{llmbox}
\end{figure*}
\begin{figure*}[hbtp]
    \centering
\begin{userbox}[Reward Calculator]
\begin{verbatim}
Ground Truth: astropy/modeling/separable.py::_cstack
Function Recall: 1.0
Function Precision: 1.0
Function F1: 1.0
Function IoU: 1.0
File Recall: 1.0
File Precision: 1.0
File F1: 1.0
File IoU: 1.0
\end{verbatim}
\end{userbox}
\end{figure*}
% \begin{figure}
% \centering
% \includegraphics[width=\textwidth]{case_study.pdf}
% \caption{Illustration of the tool-calling process of \textit{astropy\_astropy\_12907} in SWE-bench\_Verified.}
% \label{fig:case_study}
% \end{figure}

%%%%%%%%%%%%%%%%%%%%%%%%%%%%%%%%%%%%%%%%%%%%%%%%%%%%%%%%%%%%%%%%%%%%%%%%%%%%%%%
%%%%%%%%%%%%%%%%%%%%%%%%%%%%%%%%%%%%%%%%%%%%%%%%%%%%%%%%%%%%%%%%%%%%%%%%%%%%%%%

\end{document}